\documentclass[12pt]{article}
\def\baselinestretch{0.98}
\usepackage{cite}
\usepackage{multirow}
\usepackage{amssymb, amsmath, amsopn, amsthm}
\usepackage{bm,amsfonts,array,calc,rotating}
\usepackage{epsfig}
\usepackage{graphicx}
\usepackage{color}
\usepackage{epic}

\definecolor{orange}{cmyk}{0,0.5,1,0}

\newcommand{\ra}{\rightarrow}
\usepackage[T1]{fontenc}

\numberwithin{equation}{section}

\long\def\@makecaption#1#2{%
  \vskip\abovecaptionskip
  \sbox\@tempboxa{{\bf #1:} #2}%
  \ifdim \wd\@tempboxa >\hsize
    {\small\bf #1:} {\small #2}\par
  \else
    \global \@minipagefalse
    \hb@xt@\hsize{\hfil\box\@tempboxa\hfil}%
  \fi
  \vskip\belowcaptionskip}

\font\cmss=cmss10 \font\cmsss=cmss10 at 7pt
\def\IZ{\relax\ifmmode\mathchoice
{\hbox{\cmss Z\kern-.4em Z}}{\hbox{\cmss Z\kern-.4em Z}}
{\lower.9pt\hbox{\cmsss Z\kern-.4em Z}} {\lower1.2pt\hbox{\cmsss
Z\kern-.4em Z}}\else{\cmss Z\kern-.4em Z}\fi}

\def\sqr#1#2{{\vcenter{\vbox{\hrule height.#2pt
 \hbox{\vrule width.#2pt height#1pt \kern#1pt
 \vrule width.#2pt}\hrule height.#2pt}}}}

\begin{document}

\begin{flushright}
\baselineskip=12pt \normalsize
{MIFPA-10-47}\\
{TUW-10-15}\\
\smallskip
\end{flushright}

\begin{center}
\Large {\textbf{On F-theory $E_6$ GUTs}} \\[2cm]
\normalsize Ching-Ming Chen$^{\sharp}$\footnote{\tt
cmchen@hep.itp.tuwien.ac.at} and Yu-Chieh
Chung$^{\natural}$\footnote{\tt ycchung@physics.tamu.edu}
\\[.25in]
\textit{$^{\sharp}$Institute for Theoretical Physics, Vienna
University of Technology\\
Wiedner Hauptstrasse 8-10, A-1040 Vienna, AUSTRIA} \\[0.25in]
\textit{$^{\natural}$Department of Physics $\&$ Astronomy, Texas A$\&$M University\\ College Station, TX 77843, USA} \\[2.5cm]
\end{center}

\renewcommand{\baselinestretch}{1.5}
\setlength{\baselineskip}{18pt}

\begin{abstract}

We approach the Minimum Supersymmetric Standard Model (MSSM) from
an $E_6$ GUT by using the spectral cover construction and
non-abelian gauge fluxes in F-theory. We start with an $E_6$
singularity unfolded from an $E_8$ singularity and obtain $E_6$
GUTs by using an $SU(3)$ spectral cover. By turning on
$SU(2)\times U(1)^2$ gauge fluxes, we obtain a rank 5 model with
the gauge group $SU(3)\times SU(2)\times U(1)^2$. Based on the
well-studied geometric backgrounds in the literature, we
demonstrate several models and discuss their phenomenology.

\end{abstract}
\vspace{3cm}


\newpage
\setcounter{page}{1}

\setcounter{footnote}{0}

\pagenumbering{arabic}

\pagestyle{plain}

\section{Introduction}

F-theory \cite{Vafa:1996zn,Vafa:1996yn,Vafa:1996xn} is a
geometrized type IIB string theory whose background is lifted to a
twelve-dimensional manifold with an elliptic fibration. The
singularities of the elliptic fibers correspond to the gauge
groups on the seven-branes \cite{Katz:1996xe, Bershadsky:1996nh}.
Particularly, F-theory allows $E$-type singularities which
inspired the study of constructing the Grand Unification Theory
(GUT) local models admitting the down-type quark Yukawa couplings
\cite{BHV:2008I,BHV:2008II,Donagi:2008ll,Donagi:2008sl}. Recently,
F-theory and spectral cover construction
\cite{Donagi:2008ll,Donagi:global} originally introduced in the
heterotic string compactifications \cite{Friedman:1997yq} have
been used to build an $SU(5)$ GUT with an $SU(5)$ cover
\cite{Donagi:2008sl, Donagi:2008ll, Hayashi:2010zp,
Heckman:2010pv, Blumenhagen:2010ja, Jockers:2009ti, Grimm:2009yu,
Grimm:2009sy, Esole:global01, Curio:global, Collinucci:global03,
Collinucci:global02, Watari:global, Hebecker:global,
Donagi:global, Caltech:global01, Caltech:global02,
Caltech:global03, Blumenhagen:global01, Blumenhagen:global02,
Tartar:globalFlavor01, Tartar:globalFlavor02,
Other:global,other:global02, Grimm:global01, Dudas:2010zb,
King:2010mq, Dudas:2009hu}, a flipped $SU(5)$ and an $SO(10)$ with
$SU(4)$ covers \cite{Chen:2010tp, Chen:2010ts, Kuflik:2010dg}, and
an MSSM with an $SU(5)\times U(1)$ cover \cite{Choi:2010su,
Choi:2010nf}. The studies in global models can be found in
\cite{Grimm:2010ez, Marsano:2010ix, Chung:2010bn}. For a
systematic review of recent progress of F-theory compactifications
and model buildings, see \cite{Weigand:2010wm}.

To break the GUT symmetry in F-theory models, one can either use
Wilson lines \cite{BHV:2008I, Braun:2010hr} or introduce a
supersymmetric $U(1)$ flux corresponding to a fractional line
bundle \cite{BHV:2008II, Caltech:global01, Caltech:global02,
Caltech:global03, Chen:2010ts, Chen:2010tp}. In local models, an
abelian or a non-abelian flux of the rank higher than two may be
turned on on the bulk to break the gauge group \cite{BHV:2008II}.
Following this idea, an MSSM model from breaking an $SU(6)$ model
by an $U(1)\times U(1)$ gauge flux has been studied
\cite{Chung:2009ib}. There are two kinds of rank three fluxes,
$U(1)^3$ and $SU(2)\times U(1)^2$, both embedded in the $E_6$
gauge group with commutants including the Standard Model (SM)
gauge structure. We are particularly interested in the second case
containing a non-abelian $SU(2)$ gauge flux. In this paper, we
shall study the physics of the $E_6$ GUT model
\cite{Gursey:1975ki} broken by the $SU(2)\times U(1)^2$ fluxes.
$E_6$ GUT models with $U(1)_{PQ}$ symmetry in local F-theory has
been explored in \cite{Heckman:2008qt}. The detailed study of the
non-abelian fluxes and the corresponding vector bundles will be
presented elsewhere \cite{Chung:2010xx}.

There are many breaking routes from $E_6$ to a subgroup containing
the SM gauge group, such as via $SO(10)$ and then $SU(5)$, via
$SU(6)$, via Pati-Salam, or via trinification. Basically, these
breaking routes end up with two resulting gauge groups,
$G_1:SU(3)\times SU(2)_L\times U(1)^3$ and $G_2:SU(3)\times
SU(2)_L\times SU(2) \times U(1)^2$. These two subgroups are
referred to as extended MSSM models of rank 6. By suitable
rotation of the $U(1)$ gauge groups and the third component of the
$SU(2)$ gauge group, we can show that these two subgroups are
equivalent. It was found that the extended MSSM models can be
obtained from an $E_6$ unification by an $SU(2)\times U(1)^2$ or
$U(1)^3$ flux\footnote{For breaking scenarios via discrete Wilson
lines in the context of orbifold constructions, please
see\cite{Braam:2010sy} and references therein.} in the heterotic
string models \cite{Witten:1985xc}. In the literature the gauge
group obtained by breaking $E_6$ can be rank 5 or rank 6 depending
on the flux turned on \cite{Witten:1985xc, del-Aguila:1985cb,
Breit:1985ud, Dine:1985vv, Cecotti:1985by, Candelas:1985en,
Derendinger:1985kk, Ellis:1986yg, Hewett:1988xc}. When a
non-abelian flux $SU(2)\times U(1)^2$ is turned on, $E_6$ is
broken directly to a rank 5 model with a gauge group $SU(3)\times
SU(2)_L \times U(1)_Y\times U(1)_{\eta}$ after rearranging the
$U(1)$s. Normally rank 6 models have more degrees of freedom with
which to solve the problems in phenomenology. However, the $U(1)$
gauge groups induce additional gauge bosons and increase exotic
fields. By giving a large VEV to one of the $U(1)$ gauge groups,
the rank 6 models can be further reduced to the so-called
effective rank 5 models. By arranging the matter assignments, one
can build many interesting low energy models, such as $SU(3)\times
SU(2)\times U(1)_Y\times U(1)_N$. In the rank 6 model, $U(1)_N$ is
inherited from the third $U(1)$ gaining a VEV, whereas in the rank
5 model, $U(1)_{\eta}$ is fixed and does not possess additional
symmetries.

On the other hand, one of the motivations to consider models with
an additional gauge group $U(1)'$ as a gauge extension of the
Standard Model (NMSSM) is for solving the $\mu$-problem. The
minimum matter content for such a model with gauge group
$SU(3)\times SU(2)\times U(1)_Y\times U(1)'$ includes the MSSM
fermions, two Higgs doublets $H$ and $\bar H$, an SM singlet $S$
with a non-zero $U(1)'$ charge, and exotic color triplets. The
effective scale of $\mu$-term can arise from the coupling $SH\bar
H$ when the singlet $S$ acquires a VEV. The radiative breaking of
the $U(1)'$ gauge symmetry is usually achieved by the large Yukawa
couplings between the singlet $S$ and the exotic fields.  This
model can be naturally embedded in a model with the $E_6$ gauge
group while the fields mentioned above are included in the three
families of $\bf 27$-plets. For the desire of gauge unification
without introducing anomalies, a pair of Higgs-like doublets from
one or more additional $(\bf 27+\overline{\bf 27})$ is also
needed. Recently, the minimum MSSM from the $E_6$ GUT has been
studied, for example, in \cite{Langacker:1998tc, Athron:2009cb,
King:2005jy, King:2006rh, Howl:2007zi}, and phenomenology such as
the neutrino physics \cite{Ma:1995xk}, leptogenesis
\cite{Hambye:2000bn}, and baryogenesis \cite{Ma:2000jf} were also
discussed.

In this paper we construct $E_6$ GUT models in F-theory by using
the spectral cover construction and study their breaking down to
the rank 5 extended MSSM by turning on the non-abelian fluxes. We
only consider the case that the Higgs multiplets are located on a
different $\bf 27$ due to the reasons of desiring for more degrees
of freedom as well as the singularity structure of Yukawa coupling
in F-theory. We represent a few examples corresponding to two
spectral cover factorizations. In the example of $(2,1)$
factorization in $dP_7$, all the fermions are located on one $\bf
27$ curve and the introduction of fluxes for gauge breaking
results in extra copies of quarks and leptons which are exotic to
the conventional three-generation $E_6$ models. We find a better
model in the $(1,1,1)$ factorization where the fermions are from
two different $\bf 27$ curves and there is only a pair of
vector-like triplet exotic field. Both examples in $dP_7$ contain
exotic fields on the Higgs $\bf 27$ curve, and we assume they
obtain zero vacuum expectation values.

The organization of the rest of the paper is as follows: in
section 2, we give a brief review of the $SU(3)$ spectral cover
and its factorizations. In section 3, we discuss the subgroups of
$E_6$ and introduce non-abelian $SU(2)\times U(1)^2$ fluxes.
Tadpole cancellation conditions for the model building are
discussed in section 4. We demonstrate several numerical results
of rank 5 models in section 5, and then conclude in the last
section.

\section{Spectral Cover}

In this section we briefly review the construction of an $SU(3)$
spectral cover inducing an $SU(3)$ Higgs bundle breaking the gauge
group $E_8$ down to $E_6$. We also construct $(2,1)$ and $(1,1,1)$
factorizations of the cover as well as universal fluxes for
semi-local model building.

\subsection{$SU(3)$ Spectral Cover}

Let $X_4$ be an elliptically fibered Calabi-Yau fourfold
$\pi_{X_4}:X_4\ra B_3$ with a section $\sigma_{B_3}:B_3\ra X_4$
and $S$ be one component of the discriminant locus of $X_4$ with a
projection $\widetilde{\pi}: X_4\ra S$, where $X_4$ develops an
$E_6$ singularity\footnote{From now on, $S$ will be assumed to be
a del Pezzo surface unless otherwise stated
\cite{delPezzo:01,delPezzo:02}.}. To describe $X_4$, let us
consider the following Tate model \cite{Bershadsky:1996nh}:
\begin{equation}
y^2=x^3+\mathbf{b}_3yz^2+\mathbf{b}_2xz^3+\mathbf{b}_0z^5,\label{Tate
model E_6}
\end{equation}
where $x$, $y$ are the coordinates of the fibration and $z$ is the
coordinate of the normal direction of $S$ in $B_3$. Note that the
coefficients $\mathbf{b}_k$ generically depend on the coordinate
$z$ and that Eq.~(\ref{Tate model E_6}) can be regarded as
unfolding of an $E_8$ singularity\footnote{If
$\mathbf{b}_3=\mathbf{b}_2=0$, the elliptic fibration
$y^2=x^3+\mathbf{b}_0z^5$ possess an $E_8$ singularity at z=0. }
into an $E_6$ singularity. For convenience, we define the
shorthand notations $c_1(S)\equiv c_1$, $t\equiv -c_1(N_{S/B_3})$,
and $\eta\equiv 6c_1-t$ where $c_1$ is the first Chern class and
$N_{S/B_3}$ is the normal bundle of $S$ in $B_3$. To maintain the
Calabi-Yau condition $c_1(X_4)=0$, it is required that $x$ and $y$
in Eq.~(\ref{Tate model E_6}) are sections of $K_{B_3}^{-4}$ and
$K_{B_3}^{-6}$, respectively. It follows that the homological
classes $[\mathbf{b}_k]$ are $\eta-kc_1$. Note that the fiber
$\widetilde{\pi}^{-1}(b)$ for $b\in S$ is an ALE space
\cite{Math1,Math2,Math3,Math4,Douglas:1996sw,Math5}. The
singularity of the fiber over $S$ is determined by the volumes
$\lambda_k$ of $(-2)$ 2-cycles of the ALE space. So unfolding a
singularity corresponds to setting the volumes of some of these
2-cycles finite. In the Tate model Eq. $(\ref{Tate model E_6})$,
the fibration singularity is determined by the coefficients
$\mathbf{b}_k$. Indeed the coefficients $\mathbf{b}_k$ encode the
information of the volumes $\lambda_k$. In what follows, we shall
introduce the spectral cover construction making the relation
between the coefficients $\mathbf{b}_k$ in Eq. $(\ref{Tate model
E_6})$ and the volumes $\lambda_k$ of $(-2)$ 2-cycles
manifest\footnote{For more details, please see
\cite{Donagi:global} and references therein.}. Before introducing
the spectral cover, we would like to briefly review the BPS
equations arising from the compactification of the
eight-dimensional $\mathcal{N}=1$ super-Yang-Mills theory on $S$.
The details could be found in
\cite{BHV:2008I,Donagi:2008ll,Tartar:globalFlavor01}.

Let us consider the eight-dimensional $\mathcal{N}=1$ gauge theory
compactified on $S$. To obtain unbroken $\mathcal{N}=1$
supersymmetry in four dimensions, it was shown that the bosonic
fields, a gauge connection $A$ and an adjoint Higgs field $\Phi$,
have to satisfy the following BPS equations:
\begin{equation}
\left\{\begin{array}{l} F_A\wedge \omega_S+\frac{i}{2}[\Phi^{\dagger},\Phi]=0\\
F_A^{2,0}=F_A^{0,2}=0\\
\bar\partial_A\Phi=0, \label{BPS}
\end{array}   \right.
\end{equation}
where $F$ is the curvature two-form of $A$ and $\omega_S$ is a
K\"ahler form of $S$. To solve BPS equations, one may take $V$ as
a holomorphic vector bundle over $S$ with the connection $A$ and
$\Phi$ being holomorphic. The simplest solution for $(A,\Phi)$ is
that $\Phi$ is diagonal and $V$ is a stable bundle. In particular,
let us consider a $3\times 3$ case as follows:
\begin{eqnarray}
\Phi=\left(\begin{array}{@{}ccc@{}} \lambda_1 & 0 & 0
\\0 & \lambda_2 & 0 \\0& 0
&\lambda_3
\end{array}\right),~~~~~~~~~~\sum_{k=1}^{3}\lambda_k=0,\label{Higgs Field 3x3}
\end{eqnarray}
where $\lambda_k$ is holomorphic for $k=1,2,3$. In this case
$[\Phi^{\dagger},\Phi]=0$ and Eq.~(\ref{BPS}) is then reduced to
the Hermitian Yang-Mills equations
\begin{equation}
F_A^{2,0}=F_A^{0,2}=0,\;\;\;\;\;\;\;\;\;\;\;\;\;\;\;\;F_A\wedge\omega_S=0.\label{HYM00}
\end{equation}
The low energy spectrum is therefore decoupled to $\Phi$ and only
depends on the Hermitian Yang-Mills connection $A$. The
eigenvalues $\lambda_k$ characterize the locations of intersecting
seven-branes. Alternatively, the information of intersecting
seven-branes can be encoded in the characteristic polynomial
$P_{\Phi}(s)={\rm det}(sI-\Phi)$ associated with a spectral cover
over $S$. For generically diagonal $\Phi$, the polynomial equation
$P_{\Phi}(s)=0$ has distinct roots and the associated spectral
cover is smooth. However, it is not the case when $\Phi$ is upper
triangular in the following form
\begin{eqnarray}
\Phi=\left(\begin{array}{@{}ccc@{}} 0 & a & b
\\0 & 0 & c
\\0 & 0 & 0
\end{array}\right).\label{Nilp01}
\end{eqnarray}
In this case $P_{\Phi}(s)=0$ is singular and the spectrum is
coupled to $\Phi$ due to $[\Phi^{\dagger},\Phi]\neq 0$. Moreover,
the polynomial $P_{\Phi}(s)$ may not capture the entire
information of the system any more. In particular, one has to
specify not only the spectral polynomial $P_{\Phi}(s)$ but also
the Higgs field $\Phi$ to calculate the spectrum. Such
configurations of seven-branes characterized by upper triangular
$\Phi$ are called $T$-branes. For the detailed analysis of
$T$-branes, we refer readers to \cite{Cecotti:2010bp}. In what
follows, we shall focus on the case of Eq.~(\ref{Higgs Field 3x3})
and its associated spectral cover. Notice that the polynomial
equation
\begin{equation}
b_0{\det}(sI-\Phi)=b_0s^3+b_2s+b_3=0\label{Affine SU3 cover}
\end{equation}
defines a three-sheeted cover of $S$ inside the total space of the
canonical bundle $K_S\ra S$, a local Calabi-Yau threefold, where
$b_k\equiv \mathbf{b}_k|_{z=0},\;k=0,2,3$. However, this threefold
is non-compact. For well-defined intersection numbers, one can
compactify the non-compact threefold to the total space of
projective bundle $\mathbb{P}(\mathcal{O}_S\oplus K_S)$ over $S$.
Let us define $X$ the total space of the projective bundle with
two sections $U$, $V$ and with a projection map $\pi:X\ra S$. The
homological classes of zero sections $\{U=0\}$ and $\{V=0\}$ are
$\sigma$ and $\sigma+c_1$, respectively. In compact threefold $X$,
the spectral cover Eq.~(\ref{Affine SU3 cover}) can be expressed
as a homogeneous polynomial as follows:
\begin{equation}
\mathcal{C}^{(3)}:\;b_0U^3+b_2UV^2+b_3V^3\equiv
b_0\prod_{k=1}^3(U+\lambda_k V)=0,\label{SU3 cover}
\end{equation}
with a projection map $p_3:\mathcal{C}^{(3)}\ra S$. The
homological class of $\mathcal{C}^{(3)}$ is given by
$[\mathcal{C}^{(3)}]=3\sigma+\pi^{\ast}\eta$. The singularities
get enhanced at some loci of $S$. Let us consider the following
breaking pattern
\begin{eqnarray}
\begin{array}{ccl}
E_8 & \xrightarrow & E_6\times SU(3)\\ {\bf 248} & \xrightarrow &
{\bf (78,1)}+{\bf (1,8)}+{\bf (27,3)}+{\bf (\overline{27},\bar
3)}.
\end{array}
\end{eqnarray}
The matter ${\bf 27}$ is localized on the curve $\Sigma_{\bf 27}$
given by the locus of $\{b_3=0\}$ where the singularity $E_6$ is
enhanced to $E_7$, so it implies the homological class of
$[\Sigma_{\bf 27}]$ is $\eta-3c_1$ in $S$. Alternatively, it
follows from $\lambda_i=0$ in Eq.~(\ref{SU3 cover}) that the
homological class of $[\Sigma_{\bf 27}]$ can be also computed by
$[\mathcal{C}^{(3)}]\cdot \sigma|_{\sigma}=\eta-3c_1$. With a
spectral cover $\mathcal{C}^{(3)}$, one can obtain a Higgs bundle
$p_{3\ast}\mathcal{L}$ on $S$ by the pushforward of a line bundle
$\mathcal{L}$ on $\mathcal{C}^{(3)}$. To maintain the traceless
condition $c_1(p_{3\ast}\mathcal{L})=0$, it is required that
$p_{3\ast}\gamma^{(3)}=0$ where $c_1(\mathcal{L})\equiv
\gamma^{(3)}+\frac{1}{2}r^{(3)}\in H_4(X,\mathbb{Z})$ and
$r^{(3)}$ is the ramification divisor of the projection map
$p_3:\mathcal{C}^{(3)}\ra S$. Up to a constant, the unique
solution of the traceless condition $p_{3\ast}\gamma^{(3)}=0$ is
$\gamma^{(3)}=(3-p^{\ast}_3p_{3\ast})[\mathcal{C}^{(3)}]\cdot\sigma$,
and one can calculate the chiral spectrum by turning on the
traceless flux $\gamma^{(3)}$. More precisely, the net chirality
$N_{\bf 27}$ of the matter field ${\bf 27}$ can be computed as
\begin{equation}
N_{\bf 27}=\gamma^{(3)}\cdot\Sigma_{\bf
27}=-\eta\cdot_S(\eta-3c_1).\label{Chirality 27}
\end{equation}
To obtain three generations for ${\bf 27}$, it is required that
$(6c_1-t)\cdot_S(3c_1-t)=-3$ which is a non-trivial constraint on
embedding of $S$ into the Calabi-Yau fourfold $X_4$. On the other
hand, the irreducible cover $\mathcal{C}^{(3)}$ only provides a
single matter curve, so we need more matter curves and more
degrees of freedom on the cover flux to promise realistic models.
Therefore we shall study the factorizations of the spectral cover
$\mathcal{C}^{(3)}$ in what follows.

\subsection{(2,1) Factorization}

Let us consider the factorization $\mathcal{C}^{(3)}\ra
\mathcal{C}^{(a)}\times \mathcal{C}^{(b)}$:
\begin{equation}
b_0U^3+b_2UV^2+b_3V^3=(a_0U^2+a_1UV+a_2V^2)(d_0U+d_1V)\label{(2,1)
factorization}
\end{equation}
with projection maps $p_a:\mathcal{C}^{(a)}\ra S$ and
$p_b:\mathcal{C}^{(b)}\ra S$, respectively. Let $[d_1]\equiv\xi$,
one can write the homological class of remaining sections as
\begin{equation}
[a_n]=\eta-(n+1)c_1-\xi,\;n=0,1,2,\;\;\;\;\;[d_0]=c_1+\xi.\label{sections
in (2,1) factorization}
\end{equation}
It follows from Eqs.~(\ref{(2,1) factorization}) and
(\ref{sections in (2,1) factorization}) that the homological
classes of the covers $\mathcal{C}^{(a)}$ and $\mathcal{C}^{(b)}$
are given by
\begin{equation}
[\mathcal{C}^{(a)}]=2\sigma+\pi^{\ast}(\eta-\xi-c_1),
~~~~[\mathcal{C}^{(b)}]=\sigma+\pi^{\ast}(\xi+c_1).\label{class of
(2,1) factorization}
\end{equation}
With the homological classes $[\mathcal{C}^{(a)}]$ and
$[\mathcal{C}^{(b)}]$, one can compute the homological classes of
matter curves $\Sigma_{\bf 27}^{(a)}$ and $\Sigma_{\bf 27}^{(b)}$
as
\begin{equation}
[\Sigma_{\bf
27}^{(a)}]=[\mathcal{C}^{(a)}]\cdot\sigma|_{\sigma}=\eta-3c_1-\xi,~~~[\Sigma_{\bf
27}^{(b)}]=[\mathcal{C}^{(b)}]\cdot\sigma|_{\sigma}=\xi.\label{class
of matter curves in (2,1) factorization}
\end{equation}
The ramification divisors of the maps $p_a:\mathcal{C}^{(a)}\ra S$
and $p_b:\mathcal{C}^{(b)}\ra S$ are given by
\begin{equation}
r^{(a)}=[\mathcal{C}^{(a)}]\cdot\pi^{\ast}(\eta-2c_1-\xi),~~
r^{(b)}=[\mathcal{C}^{(b)}]\cdot(-\sigma+ \pi^{\ast}\xi).
\end{equation}
The traceless fluxes $\gamma^{(a)}_0$ and $\gamma^{(b)}_0$ is
defined as $(2-p^{\ast}_{a}p_{a\ast})[\mathcal{C}^{(a)}]\cdot
\sigma$ and $(1-p^{\ast}_{b}p_{b\ast})[\mathcal{C}^{(b)}]\cdot
\sigma$, respectively, where $p_{a\ast}\gamma^{(a)}_0=0$ and
$p_{b\ast}\gamma^{(b)}_0=0$. The explicit forms of the traceless
fluxes $\gamma^{(a)}_0$ and $\gamma^{(b)}_0$ are given by
\begin{equation}
\gamma^{(a)}_0 =[\mathcal{C}^{(a)}]\cdot(2\sigma- \pi^{\ast}
(\eta-3c_1-\xi)),~~~ \gamma^{(b)}_0= [\mathcal{C}^{(b)}]\cdot
(\sigma-\pi^{\ast}\xi).
\end{equation}
The chirality of matter ${\bf 27}$ on each matter curve due to the
fluxes $\gamma^{(a)}_0$ and $\gamma^{(b)}_0$ is then shown in
Table \ref{Induced Chirality gamma 2,1}.
\begin{table}[h]
\begin{center}
\renewcommand{\arraystretch}{.9}
\begin{tabular}{|c|c|c|c|} \hline
 & $\gamma_0^{(a)}$ & $\gamma_0^{(b)}$ \\ \hline
 ${\bf 27}^{(a)}$ & $-(\eta-c_1-\xi)\cdot_S(\eta-3c_1-\xi)$ & $0$ \\ \hline
 ${\bf 27}^{(b)}$ &$0$ & $-\xi\cdot_S (c_1+\xi)$  \\ \hline
\end{tabular}
\caption{Chirality induced by the fluxes $\gamma^{(a)}_0$ and
$\gamma^{(b)}_0$.} \label{Induced Chirality gamma 2,1}
\end{center}
\end{table}

Due to the factorization, one can introduce the additional fluxes
$\delta^{(a)}=(1-p^{\ast}_{b}p_{a\ast})[\mathcal{C}^{(a)}]\cdot
\sigma$ and
$\delta^{(b)}=(2-p^{\ast}_{a}p_{b\ast})[\mathcal{C}^{(b)}]\cdot
\sigma$. It is not difficult to obtain  \cite{Caltech:global03}:
\begin{eqnarray}
\delta^{(a)} =[\mathcal{C}^{(a)}]\cdot \sigma
-[\mathcal{C}^{(b)}]\cdot\pi^{\ast} (\eta-3c_1-\xi),~~~
\delta^{(b)}= [\mathcal{C}^{(b)}]\cdot
2\sigma-[\mathcal{C}^{(a)}]\cdot \pi^{\ast}\xi.
\end{eqnarray}
Also for any $\rho \in H_2(S,\mathbb{R})$, one can define a
non-trivial flux $\tilde{\rho}$ as
\begin{equation}
\tilde{\rho}=(2p_b^{\ast}-p_a^{\ast}) \rho,
\end{equation}
then the chirality induced by these additional fluxes on each
matter curve is summarized in Table {\ref{Induced Chirality delta
2,1}}.
\begin{table}[h]
\begin{center}
\renewcommand{\arraystretch}{.9}
\begin{tabular}{|c|c|c|c|} \hline
 & $\delta^{(a)}$ & $\delta^{(b)}$ & $\tilde{\rho}$ \\ \hline
 ${\bf 27}^{(a)}$ & $-c_1\cdot_S(\eta-3c_1-\xi)$ & $-\xi\cdot_S(\eta-3c_1-\xi)$ & $-\rho\cdot_S(\eta-3c_1-\xi)$ \\ \hline
 ${\bf 27}^{(b)}$ & $-\xi\cdot_S(\eta-3c_1-\xi)$ & $-2c_1\cdot_S\xi$ & $2\rho\cdot_S\xi$  \\ \hline
\end{tabular}
\caption{Chirality induced by the fluxes $\delta^{(a)}$,
$\delta^{(b)}$, and $\tilde{\rho}$.}\label{Induced Chirality delta
2,1}
\end{center}
\end{table}

The total flux $\Gamma$ is then a linear combination of the fluxes
above:
\begin{equation}
\Gamma=k_a \gamma_0^{(a)} +k_b \gamma_0^{(b)} + m_a \delta^{(a)}+
m_b \delta^{(b)} +\tilde{\rho}\equiv\Gamma^{(a)}+\Gamma^{(b)},
\end{equation}
where
\begin{eqnarray}
\Gamma^{(a)} \equiv [\mathcal{C}^{(a)}]\cdot
[\tilde{\mathcal{C}}^{(a)}] =[\mathcal{C}^{(a)}]\cdot [
(2k_a+m_a)\sigma
-\pi^{\ast}( k_a(\eta-3c_1-\xi) +m_b\xi +\rho ) ],~~ \\
\Gamma^{(b)} \equiv [\mathcal{C}^{(b)}]\cdot
[\tilde{\mathcal{C}}^{(b)}] = [\mathcal{C}^{(b)}]\cdot [
(k_b+2m_b)\sigma -\pi^{\ast}( k_b\xi +m_a(\eta-3c_1-\xi) -2\rho )
].~~\,
\end{eqnarray}
The parameters $k_a$, $k_b$, $m_a$, $m_b$ will be determined later
by the physical and consistency conditions. In addition, by
\begin{eqnarray}
&&p_{a\ast}\Gamma^{(a)} =  m_a(\eta-3c_1-\xi) -2m_b\xi -2\rho, \\
&&p_{b\ast}\Gamma^{(b)} =  -m_a(\eta-3c_1-\xi) +2m_b\xi +2\rho,
\end{eqnarray}
we find that $\Gamma^{(a)}$ and $\Gamma^{(a)}$ indeed satisfy the
traceless condition
$p_{a\ast}\Gamma^{(a)}+p_{b\ast}\Gamma^{(b)}=0$. In the $(2,1)$
factorization, the quantization conditions are then given by
\begin{eqnarray}
(2k_a+m_a)\sigma -\pi^{\ast}( k_a(\eta-3c_1-\xi) +m_b\xi
+\rho -\frac{1}{2}(\eta-2c_1-\xi) ) \in H_4(X,\mathbb{Z}),\\
(k_b+2m_b-\frac{1}{2})\sigma -\pi^{\ast}( k_b\xi
+m_a(\eta-3c_1-\xi) -2\rho -\frac{1}{2}\xi ) \in H_4(
X,\mathbb{Z}).
\end{eqnarray}
In addition, the supersymmetry condition is
\begin{equation}
[m_a(\eta-3c_1-\xi) -2m_b\xi -2\rho]\cdot_S[\omega]=0,
\end{equation}
where $[\omega]$ is an ample divisor dual to a K\"ahler form of
$S$.

\subsection{(1,1,1) Factorization}

Let us consider the factorization $\mathcal{C}^{(3)}\ra
\mathcal{C}^{(l_1)}\times \mathcal{C}^{(l_2)}\times
\mathcal{C}^{(l_3)}$:
\begin{equation}
b_0U^3+b_2UV^2+b_3V^3=(f_0U+f_1V)(g_0U+g_1V)(h_0U+h_1V),\label{(1,1,1)
factorization}
\end{equation}
with the projection maps $p_{l_1}:\mathcal{C}^{(l_1)}\ra S$,
$p_{l_2}:\mathcal{C}^{(l_2)}\ra S$, and
$p_{l_3}:\mathcal{C}^{(l_3)}\ra S$. Let $[g_1]\equiv\xi_1$ and
$[h_1]\equiv\xi_2$, the homological classes of the remaining
sections are
\begin{equation}
[f_m]=\eta-(m+2)c_1-\xi_1-\xi_2,\;m=0,1.\;\;\;[g_0]=c_1+\xi_1,\;[h_0]=c_1+\xi_2.\label{sections
in (1,1,1) factorization}
\end{equation}
It follows from Eqs.~(\ref{(1,1,1) factorization}) and
(\ref{sections in (1,1,1) factorization}) that the homological
classes of the covers $\mathcal{C}^{(l_1)}$,
$\mathcal{C}^{(l_2)}$, and $\mathcal{C}^{(l_3)}$ are given by
\begin{equation}
[\mathcal{C}^{(l_1)}]=\sigma+\pi^{\ast}(\eta-2c_1-\xi_1-\xi_2),
~~~[\mathcal{C}^{(l_2)}]=\sigma+\pi^{\ast}(\xi_1+c_1),
~~~[\mathcal{C}^{(l_3)}]=\sigma+\pi^{\ast}(\xi_2+c_1).
\label{class of (1,1,1) factorization}
\end{equation}
The homological classes of the matter curves can be obtained from
the intersection $[\mathcal{C}^{(l_i)}]\cdot \sigma|_{\sigma}$:
\begin{equation}
[\Sigma_{\bf 27}^{(l_1)}]=\eta-3c_1-\xi_1-\xi_2,~~~[\Sigma_{\bf
27}^{(l_2)}]=\xi_1,~~~[\Sigma_{\bf 27}^{(l_3)}]=\xi_2.
\label{class of matter curves in (1,1,1) factorization}
\end{equation}
In the $(1,1,1)$ factorization, the ramification divisors are
given by
\begin{equation}
r_{l_1}=[\mathcal{C}^{(l_1)}]\cdot[-\sigma+\pi^{\ast}(\eta-3c_1-\xi_1-\xi_2)],~~
r_{l_2}=[\mathcal{C}^{(l_2)}]\cdot(-\sigma+ \pi^{\ast}\xi_1),~~
r_{l_3}=[\mathcal{C}^{(l_3)}]\cdot(-\sigma+ \pi^{\ast}\xi_2).
\end{equation}
For general fluxes $\gamma^{(i)}=[\mathcal{C}^{(i)}]\cdot\sigma$,
we define the traceless fluxes $\gamma^{(i)}_0$ as
\begin{eqnarray}
&\gamma^{(l_1)}_0 =& (1-p_{l_1}^{\ast}p_{l_1\ast})\gamma^{(l_1)}
=[\mathcal{C}^{(l_1)}]\cdot[\sigma- \pi^{\ast}(\eta-3c_1-\xi_1-\xi_2)], \\
&\gamma^{(l_2)}_0 =& (1-p_{l_2}^{\ast}p_{l_2\ast})\gamma^{(l_2)}
=[\mathcal{C}^{(l_2)}]\cdot (\sigma-\pi^{\ast}\xi_1),\\
&\gamma^{(l_3)}_0 =& (1-p_{l_3}^{\ast}p_{l_3\ast})\gamma^{(l_3)}
=[\mathcal{C}^{(l_3)}]\cdot (\sigma-\pi^{\ast}\xi_2).
\end{eqnarray}
It is easy to see that $\gamma^{(i)}_0$ satisfies the condition
$p_{i\ast}\gamma^{(i)}_0=0$ for all $i$. The chirality induced by
the fluxes $\gamma^{(l_1)}_0$, $\gamma^{(l_2)}_0$, and
$\gamma^{(l_3)}_0$ is summarized in Table~\ref{Induced Chirality
gamma 1,1,1}.

\begin{table}[h]
\begin{center}
\renewcommand{\arraystretch}{.9}
\begin{tabular}{|c|c|c|c|} \hline
 & $\gamma_0^{(l_1)}$ & $\gamma_0^{(l_2)}$ & $\gamma_0^{(l_3)}$\\ \hline
 ${\bf 27}^{(l_1)}$ & $-(\eta-2c_1-\xi_1-\xi_2)\cdot_S(\eta-3c_1-\xi_1-\xi_2)$ & $0$ & $0$ \\ \hline
 ${\bf 27}^{(l_2)}$ &$0$ & $-\xi_1\cdot_S (c_1+\xi_1)$ & $0$  \\ \hline
 ${\bf 27}^{(l_3)}$ &$0$ & $0$ & $-\xi_2\cdot_S (c_1+\xi_2)$   \\ \hline
\end{tabular}
\caption{Chirality induced by the fluxes $\gamma^{(l_1)}_0$,
$\gamma^{(l_2)}_0$, and $\gamma^{(l_3)}_0$.}\label{Induced
Chirality gamma 1,1,1}
\end{center}
\end{table}

There are many choices of the additional fluxes, for simplicity,
we consider
\begin{eqnarray}
&\delta^{(l_1)}&
=[(1-p_{l_2}^{\ast}p_{l_1\ast})+(1-p_{l_3}^{\ast}p_{l_1\ast})]\gamma^{(l_1)}
\nonumber \\
&&= [\mathcal{C}^{(l_1)}]\cdot 2\sigma
-([\mathcal{C}^{(l_2)}]+[\mathcal{C}^{(l_3)}]) \cdot
\pi^{\ast}(\eta-3c_1-\xi_1-\xi_2),  \\
&\delta^{(l_2)}&
=[(1-p_{l_1}^{\ast}p_{l_2\ast})+(1-p_{l_3}^{\ast}p_{l_2\ast})]\gamma^{(l_2)}
\nonumber \\
&&= [\mathcal{C}^{(l_2)}]\cdot 2\sigma-[\mathcal{C}^{(l_1)}]\cdot
\pi^{\ast}\xi_1-[\mathcal{C}^{(l_3)}]\cdot \pi^{\ast}\xi_1, \\
&\delta^{(l_3)}&
=[(1-p_{l_1}^{\ast}p_{l_3\ast})+(1-p_{l_2}^{\ast}p_{l_3\ast})]\gamma^{(l_3)}
\nonumber \\
&&= [\mathcal{C}^{(l_3)}]\cdot 2\sigma-[\mathcal{C}^{(l_1)}]\cdot
\pi^{\ast}\xi_2-[\mathcal{C}^{(l_2)}]\cdot \pi^{\ast}\xi_2.\\
&\widehat{\rho}&=(p_{l_2}^{\ast}-p_{l_1}^{\ast})\rho_1+(p_{l_3}^{\ast}-p_{l_2}^{\ast})\rho_2
+(p_{l_1}^{\ast}-p_{l_3}^{\ast})\rho_3,
\end{eqnarray}
where $\rho_i \in H_2(S,\mathbb{R}),\;\forall i$. The chirality
induced by these additional fluxes on each matter curve is
summarized in Table \ref{Induced Chirality delta 1,1,1}.
\begin{table}[h]
\begin{center}
\renewcommand{\arraystretch}{.75}
\begin{tabular}{|c|c|c|c|c|} \hline
 & $\delta^{(l_1)}$ & $\delta^{(l_2)}$ & $\delta^{(l_3)}$ & $\widehat{\rho}$ \\ \hline
 ${\bf 27}^{(l_1)}$ & $-2c_1\cdot_S[f_1]$ & $-\xi_1\cdot_S[f_1]$ & $-\xi_2\cdot_S[f_1]$ & $(\rho_3-\rho_1)\cdot_S[f_1]$ \\ \hline
 ${\bf 27}^{(l_2)}$ & $-\xi_1\cdot_S[f_1]$ & $-2c_1\cdot_S\xi_1$ & $-\xi_1\cdot_S\xi_2$ & $(\rho_1-\rho_2)\cdot_S\xi_1$ \\ \hline
 ${\bf 27}^{(l_3)}$ & $-\xi_2\cdot_S[f_1]$ & $-\xi_1\cdot_S\xi_2$ & $-2c_1\cdot_S\xi_2$ & $(\rho_2-\rho_3)\cdot_S\xi_2$ \\ \hline
\end{tabular}
\caption{Chirality induced by the fluxes $\delta^{(l_1)}$,
$\delta^{(l_2)}$, $\delta^{(l_3)}$ and
$\widehat{\rho}$.}\label{Induced Chirality delta 1,1,1}
\end{center}
\end{table}

The total flux $\Gamma$ with the parameters $k_{l_1}$, $k_{l_2}$
$k_{l_3}$, $m_{l_1}$, $m_{l_2}$, and $m_{l_3}$ is
\cite{Caltech:global03}
\begin{equation}
\Gamma=k_{l_1} \gamma_0^{(l_1)} +k_{l_2} \gamma_0^{(l_2)}+k_{l_3}
\gamma_0^{(l_3)} + m_{l_1} \delta^{(l_1)}+ m_{l_2} \delta^{(l_2)}+
m_{l_3} \delta^{(l_3)}
+\widehat{\rho}\equiv\Gamma^{(l_1)}+\Gamma^{(l_2)}+\Gamma^{(l_3)},
\end{equation}
where
\begin{eqnarray}
\Gamma^{(l_1)} \equiv [\mathcal{C}^{(l_1)}]\cdot
[\tilde{\mathcal{C}}^{(l_1)}] =[\mathcal{C}^{(l_1)}]\cdot [
(k_{l_1}+2m_{l_1})\sigma
-\pi^{\ast}( k_{l_1}[f_1] +m_{l_2}\xi_1 +m_{l_3}\xi_2 +\rho_1-\rho_3 ) ],~~ \\
\Gamma^{(l_2)} \equiv [\mathcal{C}^{(l_2)}]\cdot
[\tilde{\mathcal{C}}^{(l_2)}] = [\mathcal{C}^{(l_2)}]\cdot [
(k_{l_2}+2m_{l_2})\sigma -\pi^{\ast}( m_{l_1}[f_1] +k_{l_2}\xi_1
+m_{l_3}\xi_2 +\rho_2-\rho_1 )
],~~ \\
\Gamma^{(l_3)} \equiv [\mathcal{C}^{(l_3)}]\cdot
[\tilde{\mathcal{C}}^{(l_3)}] = [\mathcal{C}^{(l_3)}]\cdot [
(k_{l_3}+2m_{l_3})\sigma -\pi^{\ast}( m_{l_1}[f_1] +m_{l_2}\xi_1
+k_{l_3}\xi_2 +\rho_3-\rho_2 ) ].~~
\end{eqnarray}
It is then straightforward to compute
\begin{eqnarray}
&&p_{l_1\ast}\Gamma^{(l_1)} =  2m_{l_1}(\eta-3c_1-\xi_1-\xi_2) -m_{l_2}\xi_1 -m_{l_3}\xi_2 -\rho_1+\rho_3, \\
&&p_{l_2\ast}\Gamma^{(l_2)} =  -m_{l_1}(\eta-3c_1-\xi_1-\xi_2) +2m_{l_2}\xi_1 -m_{l_3}\xi_2 -\rho_2+\rho_1, \\
&&p_{l_3\ast}\Gamma^{(l_3)} =  -m_{l_1}(\eta-3c_1-\xi_1-\xi_2)
-m_{l_2}\xi_1 +2m_{l_3}\xi_2 -\rho_3+ \rho_2.
\end{eqnarray}
The sum is zero, as it should be for the traceless condition. In
this case, the quantization conditions are given by
\begin{eqnarray}
(k_{l_1}+2m_{l_1}-\frac{1}{2})\sigma -\pi^{\ast}\{
(k_{l_1}-\frac{1}{2})[f_1] +m_{l_2}\xi_1
+m_{l_3}\xi_2+\rho_1-\rho_3 \} \in H_4(X,\mathbb{Z}),\\
(k_{l_2}+2m_{l_2}-\frac{1}{2})\sigma -\pi^{\ast}\{ m_{l_1}[f_1] +
(k_{l_2}-\frac{1}{2})\xi_1+m_{l_3}\xi_2 +\rho_2-\rho_1\} \in H_4(
X,\mathbb{Z}), \\
(k_{l_3}+2m_{l_3}-\frac{1}{2})\sigma -\pi^{\ast}\{ m_{l_1}[f_1]
+m_{l_2}\xi_1+ (k_{l_3}-\frac{1}{2})\xi_2 +\rho_3-\rho_2\} \in
H_4(X,\mathbb{Z}),
\end{eqnarray}
and the supersymmetry conditions are as follows:
\begin{eqnarray}
&&[2m_{l_1}(\eta-3c_1-\xi_1-\xi_2) -m_{l_2}\xi_1 -m_{l_3}\xi_2 -\rho_1+\rho_3]\cdot_S[\omega]=0, \\
&&[-m_{l_1}(\eta-3c_1-\xi_1-\xi_2) +2m_{l_2}\xi_1 -m_{l_3}\xi_2 -\rho_2+\rho_1]\cdot_S[\omega]=0, \\
&&[-m_{l_1}(\eta-3c_1-\xi_1-\xi_2) -m_{l_2}\xi_1 +2m_{l_3}\xi_2
-\rho_3+ \rho_2]\cdot_S[\omega]=0.
\end{eqnarray}


\section{Breaking $E_6$}

The MSSM fermion and electroweak Higgs fields can be included in
the same $\bf 27$ multiplet of a three-family $E_6$ GUT model. On
the other hand, it is possible to assign the Higgs fields to a
different ${\bf 27}_H$ multiplet where only the Higgs doublets and
singlets obtain the electroweak scale energy. The Yukawa coupling
for these two cases can be written as
\begin{equation}
\mathcal{W}\supset {\bf 27} \cdot{\bf 27} \cdot {\bf
27}~{\rm(Case~A)} ~~{\rm or}~~ {\bf 27} \cdot{\bf 27} \cdot {\bf
27}_H ~{\rm(Case~B)}.
\end{equation}
The Yukawa coupling of Case A is either a triple-intersection of
one $\bf 27$ curve or an intersection of three different curves in
F-theory model building. It is difficult to obtain a three family
model from a single curve and the geometry of a
triple-intersection is generally complicated. On other hand, it is
not easy to achieve the mass hierarchy of the third generation in
the three-curve model. Therefor we do not consider Case A in this
paper. In case B, there are two possible constructions from
spectral cover factorizations. In the $(2,1)$ factorization, the
fermions are assigned to ${\bf 27}^{(a)}$ curve and the Higgs
fields come from the other ${\bf 27}^{(b)}$ curve. The Yukawa
coupling then turns out
\begin{equation}
\mathcal{W}^{(2,1)}\supset {\bf 27}^{(a)} \cdot{\bf 27}^{(a)}
\cdot {\bf 27}^{(b)}.
\end{equation}
In the $(1,1,1)$ factorization, the matter fields are assigned to
curve ${\bf 27}^{(a)}$ and ${\bf 27}^{(b)}$ while the Higgs fields
come from the ${\bf 27}^{(c)}$ curve. In this case the Yukawa
coupling is then
\begin{equation}
\mathcal{W}^{(1,1,1)}\supset {\bf 27}^{(a)} \cdot{\bf 27}^{(b)}
\cdot {\bf 27}^{(c)}.
\end{equation}

In order to realize the MSSM in the $E_6$ GUT models, it is useful
to study the subgroups of $E_6$. In our F-theory model building we
consider the picture that the $E_6$ gauge group is broken by the
$SU(2)\times U(1)^2$ flux on the seven-branes. This flux may tilt
the chirality of the matter on the curve after $E_6$ is broken.

\subsection{Subgroups of $E_6$}

The subgroups of $E_6$ including the Standard Model gauge group
can be denoted $E_6\supset SU(3)\times SU(2)_L\times G_c$. Here
$G_c$ marks a rank 3 group which is a product of $U(1)$ or
$SU(2)$. It has been shown (for example, \cite{Gursey:1975ki,
Dine:1985vv, Hewett:1988xc, Harada:2003sb}) that by suitable
assignments of the hypercharge of the SM and the $B-L$ symmetry,
these $E_6$ subgroups with different $G_c$ are equivalent to
different matter content arrangements. This property would be
useful for the analysis of the non-abelian fluxes of type $G_c$.
In this section we will briefly review the subgroups of $E_6$.

Let us consider the following breaking patterns of $E_6$:
\begin{eqnarray}
&(1a)& E_6\ra SO(10)\times U(1) \ra SU(5)\times U(1)^2,~~ \label{Case 1a}\\
&(1b)& E_6\ra SO(10)\times U(1) \ra SU(4)\times SU(2)\times SU(2)\times U(1),~~ \label{Case 1b}\\
&(2a)& E_6\ra SU(6)\times SU(2) \ra SU(5)\times U(1)\times SU(2),~~ \label{Case 2a} \\
&(2b)& E_6\ra SU(6)\times SU(2) \ra SU(4)\times SU(2)\times U(1)\times SU(2),~~ \label{Case 2b} \\
&(2c)& E_6\ra SU(6)\times SU(2) \ra SU(3)\times SU(3)\times U(1)\times SU(2),~~ \label{Case 2c} \\
&(3)& E_6\ra SU(3)\times SU(3)\times SU(3).\label{Case 3}
\end{eqnarray}
In all of these cases, there are two possible outcomes when $E_6$
is broken down to the subgroups containing the Standard Model
group. Case $(1a)$ turns out to be
\begin{equation}
E_6\rightarrow SU(3)\times SU(2)_L\times U(1)_{Y} \times
U(1)_{\chi}\times U(1)_{\psi},  \label{SM-SU5}
\end{equation}
and the other cases become
\begin{equation}
E_6\rightarrow SU(3)\times SU(2)\times SU(2) \times U(1)_U\times
U(1)_W. \label{SM-SU2}
\end{equation}
Note that the assignments of $U(1)_U$ and $U(1)_W$ groups of the
cases $(1b)$, $(2a)$, $(2b)$, $(2c)$ and $(3)$ are different, but
they are equivalent up to linear transformations and the details
can be found in the appendix. Take case $(3)$ as an example, the
breaking is through a trinification model, therefore we can write
\begin{equation}
E_6\supset SU(3)\times SU(2)_L\times SU(2)_{(R)} \times
U(1)_{Y_L}\times U(1)_{Y_{(R)}}.
\end{equation}
The parenthesis on $R$ in $SU(2)_{(R)}$ indicates that it has
three different assignments denoted by $SU(2)_{R}$, $SU(2)_{R'}$,
and $SU(2)_{E}$ \cite{Harada:2003sb}. The third component
$I_{3(R)}$ of $SU(2)_{(R)}$ along with the quantum numbers of
$U(1)_{Y_L}$ and $U(1)_{Y_{(R)}}$ can have a linear relation to
the quantum numbers of $U(1)_{Y}$, $U(1)_{\chi}$ and $U(1)_{\psi}$
of case $(1a)$ in (\ref{SM-SU5}), i.e.,
\begin{equation}
Y=a_1Y_L+a_2Y_{(R)}+a_3I_{3(R)},~
\chi=b_1Y_L+b_2Y_{(R)}+b_3I_{3(R)},~
\psi=c_1Y_L+c_2Y_{(R)}+c_3I_{3(R)},~
\end{equation}
where $a_i$, $b_i$ and $c_i$ are coefficients of the
transformation. These three different kinds of $SU(2)_{(R)}$
assignments also confine the three different embedding of SM
matter representations into the $SU(5)$ multiplets belonging to
$\bf 27$ of $E_6$, as well as the corresponding assignments of the
hypercharge. The three assignments of $U(1)_Y$ should be
orthogonal to the three $SU(2)_{(R)}$, respectively.

The $U(1)_{B-L}$ symmetry is conserved in SUSY $E_6$ models, which
is not difficult to see from the gauge breaking via the Pati-Salam
gauge group. $U(1)_{B-L}$ has a linear relation with $U(1)_{Y_L}$,
$U(1)_{Y_{(R)}}$, and the third component of $SU(2)_{(R)}$. There
are also three $U(1)_{B-L}$ assignments orthogonal to the three
$SU(2)_{(R)}$, respectively. For consistency with the SM
structure, $U(1)_{B-L}$ and $U(1)_Y$ are not orthogonal to the
same $SU(2)_{(R)}$. Therefore, there are totally six different
charge assignments of the SM multiplets, in other words six
different embedding of SM multiplets in $\bf 27$ of $E_6$. For the
detailed analysis, we refer readers to \cite{Harada:2003sb}.

The $E_6$ subgroups listed in Eqs.~(\ref{SM-SU5}) and
(\ref{SM-SU2}) are rank 6. In heterotic string compactifications,
$E_6$ can be broken by a non-abelian flux down to a rank 5
subgroup \cite{Witten:1985xc, Candelas:1985en, Dine:1985vv,
Breit:1985ud}:
\begin{equation}
E_6\rightarrow SU(3)\times SU(2)_L\times U(1)_{Y} \times
U(1)_{\eta}. \label{SM-rank5}
\end{equation}
This model is usually marked as the $\eta$-model. Rank 6 models
\cite{Derendinger:1985kk, del-Aguila:1985cb, Ellis:1986yg} have
more symmetries, but it is common practice to give a large VEV to
one $U(1)$ gauge group to reduce them to the so called
\textit{effective} rank 5 models. For instance, from
Eq.~(\ref{SM-SU5}) the remaining abelian gauge group
$U(1)_{\theta}$ is a reduction
\begin{equation}
U(1)_{\theta}=\cos\theta U(1)_{\chi}+\sin\theta U(1)_{\psi}.
\end{equation}
Particularly, the rank 5 $\eta$-model can be regarded as a special
case of this setup by
\begin{equation}
U(1)_{\eta} = \sqrt{\frac{3}{8}} U(1)_{\chi}-\sqrt{\frac{5}{8}}
U(1)_{\psi}.
\end{equation}
In our F-theory models, a non-abelian flux $SU(2)\times U(1)^2$ is
turned on to break the $E_6$ gauge group into $SU(3)\times
SU(2)\times U(1)^2$ taken to be the $\eta$-model. However, since
$U(1)_{\eta}$ is only determined by the two $U(1)$s while the
$SU(2)$ is integrated out, the $\eta$-model does not possess the
degrees of freedom from the mixing angle $\theta$ preserving some
symmetries such as the $B-L$ symmetry \cite{Hewett:1988xc}. The
corresponding phenomenology of the F-theory rank 5 model will
basically follow the properties of the $\eta$-model.

The particle content of the $E_6$ model we will consider is
conventional. It includes three copies of {\bf 27}-plets, each
copy includes an SM ordinary family, two Higgs-type doublets, two
SM singlets, and two exotic $SU(2)$-singlet quarks. The $\bf 27$
matter content of the $SU(3)\times SU(2)\times U(1)_Y\times
U(1)_{\eta}$ model with the corresponding charges~are
\begin{equation}
\begin{array}{cc@{~}l}
{\bf 27} & \rightarrow &  Q{\bf (3,2)}_{\frac{1}{3},2} +u^c{\bf
(\bar 3,1)}_{-\frac{4}{3},2} +e^c{\bf (1,1)}_{2,2} \\
&+& L{\bf (1,2)}_{-1,-1}
+d^c{\bf (\bar 3,1)}_{\frac{2}{3},-1}+ \nu^c{\bf (1,1)}_{0,5} \\
&+& \bar D{\bf (3,1)}_{-\frac{2}{3},-4} + \bar h{\bf (1,2)}_{1,-4}\\
&+& D{\bf (\bar 3,1)}_{\frac{2}{3},-1}+ h{\bf (1,2)}_{-1,-1} +
S{\bf (1,1)}_{0,5},
\end{array}
\end{equation}
where the first subscription denotes the $U(1)_Y$ charge and the
second indicates the $U(1)_{\eta}$ charge. The superpotential for
the ${\bf 27}\cdot {\bf 27}\cdot {\bf 27}$ coupling can be
expanded as
\begin{eqnarray}
&&\mathcal{W}=\mathcal{W}_0+\mathcal{W}_1+\mathcal{W}_2+\mathcal{W}_3 +\cdots,\\
&&\mathcal{W}_0=\lambda_1 \bar hQu^c+\lambda_2hQd^c
+\lambda_3hLe^c+\lambda_4 h\bar hS + \lambda_5D\bar DS,  \\
&&\mathcal{W}_1=\lambda_6 \bar D u^c e^c + \lambda_7  D QL +
\lambda_8 \bar
D\nu^c d^c, \\
&&\mathcal{W}_2=\lambda_9 \bar DQQ + \lambda_{10} D u^c d^c,\\
&&\mathcal{W}_3=\lambda_{11} \bar h L\nu^c.
\end{eqnarray}
To avoid the terms that may cause serious phenomenological
problems, additional symmetries such as discrete symmetry should
be considered. The exotic fields are only confined by the charge,
isospin, and hypercharge assignments while their baryon and lepton
numbers remain unspecified. By assigning baryon and lepton numbers
to $D$, it is possible to forbid some of the interactions in
$\mathcal{W}$ by the conservation of baryon and lepton numbers.
For example, if the baryon number $B(\bar D)=\frac{1}{3}$ and the
lepton number $L(\bar D)=1$, $\mathcal{W}_2=0$; if $B(\bar
D)=-\frac{2}{3}$ and $L(\bar D)=0$, then $\mathcal{W}_1=0$. In the
case $B(\bar D)=\frac{1}{3}$ and $L(\bar D)=0$, $\bar D$ is
regarded as a conventional quark able to mix with the $d$-quarks,
and then decay via flavor changing neutral currents (FCNC) or
charged currents (CC) \cite{Hewett:1988xc}. By setting $B(h,\bar
h)=L(h,\bar h)=0$ and $B(S)=L(S)=0$, $h$ and $\bar h$ are the
usual MSSM Higgs doublets, and the VEV of $S$ provides a mass for
$D$. See \cite{Hewett:1988xc} for a detailed review.

Another possibility is considering the MSSM Higgs fields coming
from a different ${\bf 27}_H$ (or $\overline{\bf 27}_H$). In this
case the exotics of the matter $\bf 27$-plet are taken as the
ordinary quarks and leptons, $B(\bar D)=\frac{1}{3}$ and $L(\bar
D)=0$, as well as $B(h,\bar h, \nu^c, S)=0$ and $L(h,\bar h,
\nu^c, S)=\pm 1$. The doublets $H_1{\bf (1,2)}_{-1,-1}$, $H_2{\bf
(1,2)}_{-1,-1}$ and $\bar H_2{\bf (1,2)}_{1,-4}$, and the singlets
$H_3{\bf (1,1)}_{0,5}$ and $H_4{\bf (1,1)}_{0,5}$ of ${\bf 27}_H$
develop VEVs so that the superpotential takes the form
\begin{eqnarray}
\mathcal{W}'&\supset& \bar H_2Qu^c+H_2Qd^c
+H_2Le^c+  H_1 he^c + \bar hhH_4  \nonumber  \\
&+& \bar H_2 h S + H_2\bar hS + \bar DDH_4 + H_1 Q D + H_3 \bar D
d^c + \bar H_2 L\nu^c +\cdots.
\end{eqnarray}
We can see the mixing terms between the ordinary fermions and
their corresponding exotic fields. These kinds of mixings allow
the exotics to decay via FCNC or CC. For example, the coupling to
$W$ of charged currents (CC) for electric charge
$Q_e=\frac{2}{3},-\frac{1}{3}$ sector can be \cite{Hewett:1988xc}
\begin{equation}
\mathcal{L}^{CC}\sim \frac{g}{\sqrt{2}} (\bar u,0)_L \left (
\begin{array}{@{}cc@{}} U^u_L & 0 \\0 & 0 \end{array} \right ) \left (
\begin{array}{@{}cc@{}} I & 0 \\0 & 0 \end{array} \right )
{U^d_L}^{\dag} \gamma_{\mu}(1-\gamma_5) \left (
\begin{array}{@{}c@{}} d \\ \bar D \end{array} \right )
W^{\mu}+h.c.~, \label{CCQ}
\end{equation}
where $U^u_L$ and $I$ are $n\times n$ matrices and $U^d_L$ is a
$2n\times 2n$ matrix for $n$ generations. $U^u_L$ and $U_L^d$ are
transformations from weak eigenstates to mass eigenstates. On the
other hand for $Q_e=0,-1$ sector, if the two components of the
doublet $h$ are $h=(N,E)$, the coupling is \cite{Hewett:1988xc}
\begin{equation}
\mathcal{L}^{CC}\sim \frac{g}{2\sqrt{2}} (\bar \nu,\bar N)_L
\gamma_{\mu}(V-A \gamma_5) \left (
\begin{array}{@{}c@{}} e \\ E \end{array} \right )
W^{\mu}+h.c.~, \label{CCL}
\end{equation}
where $V$ and $A$ are $2n\times 2n$ matrices composed of the left
and right weak-mass transformations:
\begin{equation}
V=U_L^{\nu}{U_L^e}^{\dag} +U_R^{\nu} \left (
\begin{array}{@{}cc@{}} 0 & 0 \\0 & I \end{array} \right )
{U_R^e}^{\dag},~~~
A=U_L^{\nu}{U_L^e}^{\dag} -U_R^{\nu} \left (
\begin{array}{@{}cc@{}} 0 & 0 \\0 & I \end{array} \right )
{U_R^e}^{\dag}.
\end{equation}
Therefore the CC couplings allow the decays $\bar D\rightarrow
u+W$ and $E\rightarrow \nu +W$. Similarly, for the fermions couple
to the neutral gauge bosons $Z_a^{\mu}$, $a=Y,\eta$, the couplings
can be written as \cite{Hewett:1988xc}
\begin{equation}
\mathcal{L}^{NC}\sim \sum_{i,a}(\bar f_{L,i}\gamma_u C_L^{i,a}
f_{L,i}+{\rm L \leftrightarrow R})Z_a^{\mu}, \label{FCNC}
\end{equation}
where $C_{L,R}^{i,a}= U^i_{L,R} P_{L,R}^{i,a} {U^i_{L,R}}^{\dag}$,
$P_{L,R}^{i,a}$ are coupling matrices, and $f_i$ present fermions
$u$, $d$, $\bar D$, $e$, $\dots$ etc. This will allow the decays
$\bar D\rightarrow d+Z$ and $E\rightarrow e+Z$. In addition, the
$Z_\eta$ boson can mix with $Z$ boson through the conventional
$Z-Z'$ mixing mechanism and decay into either fermion pairs, SUSY
partners, $W$ bosons, higgsinos and gauginos, or $Z$ boson with
Higgses.  More phenomenology details can be referred to
\cite{Hewett:1988xc}\footnote{We only briefly discussed the
phenomenology of the scenario that fermions and Higgs fields are
from two different $\bf 27$ multiplets (curves) due to the
F-theory construction. We also focused on the case of $h$-leptons
and $D$-quarks assignments. There could be additional conditions
from F-theory or geometry to confine the degrees of freedom left
for the exotic fields. We leave this topic for our future study.}.

There can be one or more additional Higgs-like doublets from
$({\bf 27}+\overline{\bf 27})$ vector-like pairs preserving the
gauge unification without introducing anomalies. In summary, with
the picture of electroweak Higgs fields from a different ${\bf
27}_H$, the minimum spectrum at low energy is
\begin{equation}
3\times {\bf 27}+({\bf 27}_H)+({\bf 27}+\overline{\bf 27}).
\end{equation}

\subsection{Non-abelian Gauge Fluxes}

In what follows, we shall analyze the effects on the chirality
after the $SU(2)\times U(1)^2$ flux is turned on. We choose the
breaking chain $(1b)$ in Eq.~(\ref{Case 1b}) via $SO(10)$ and
$SU(4)\times SU(2)\times SU(2)$. When the flux is turned on, the
matter on the bulk is decomposed as
\begin{equation}
\begin{array}{ccl}
E_6 & \xrightarrow[U(1)_a] & SO(10)\times [U(1)_a] \\
& \xrightarrow[SU(2)] & SU(4)\times SU(2)_1\times [SU(2)_2\times
U(1)_a] \\
& \xrightarrow[U(1)_b] & SU(3)\times SU(2)_1\times [SU(2)_2\times
U(1)_a\times U(1)_b] \\ \\
{\bf 78} & \rightarrow & {\bf 45}_0+{\bf 1}_0+{\bf 16}_{-3}+{\bf \overline{16}}_3 \\
& \rightarrow & {\bf (15,1,1)}_0+{\bf (6,2,2)}_0+{\bf (1,3,1)}_0+{\bf (1,1,3)}_0+{\bf (1,1,1)}_0\\
&& [{\bf (4,2,1)}_{-3}+{\bf (\bar 4,1,2)}_{-3}+c.c.]\\
& \rightarrow & {\bf (8,1,1)}_{0,0} + {\bf (3,1,1)}_{0,-4} +{\bf
(\bar 3,1,1)}_{0,4}
+{\bf (1,1,1)}_{0,0}  \\
&& +{\bf (3,2,2)}_{0,2}+{\bf (\bar 3,2,2)}_{0,-2}+{\bf (1,3,1)}_{0,0}+{\bf (1,1,3)}_{0,0}+{\bf (1,1,1)}_{0,0}\\
&& +[{\bf (3,2,1)}_{-3,-1}+{\bf (1,2,1)}_{-3,3} +{\bf (\bar
3,1,2)}_{-3,1} + {\bf (1,1,2)}_{-3,-3}+c.c.].\label{E_6 decomp}
\end{array}
\end{equation}
The SM hypercharge is defined as
\begin{equation}
U(1)_Y= \frac{1}{2}[U(1)_a+\frac{1}{3}U(1)_b].
\end{equation}
Under the breaking pattern (\ref{E_6 decomp}), the gauge group
$E_6$ can be broken down to $SU(3)\times SU(2)_1\times
U(1)_a\times U(1)_b$ by turning on a gauge bundle on $S$ with the
structure group $SU(2)_2\times U(1)_a\times U(1)_b$. Let us define
$L_1$ and $L_2$ to be the line bundles associated with $U(1)_a$
and $U(1)_b$, respectively. $V_2$ is defined as a vector bundle of
rank two with the structure group $SU(2)$. To preserve
supersymmtry, the connection of the gauge bundle $W=V_2\oplus
L_1\oplus L_2$ has to satisfy the Hermitian Yang-Mills equations
(\ref{HYM00})\footnote{More precisely, $L_1$ and $L_2$ are
fractional line
bundles\cite{BHV:2008I,BHV:2008II,Donagi:2008ll,Donagi:2008sl}.}.
It was shown in \cite{Donalson,UY} that the bundle $W$ has to be
poly-stable with
$\mu_{[\omega]}(V_2)=\mu_{[\omega]}(L_1)=\mu_{[\omega]}(L_2)=0$,
where slope $\mu_{[\omega]}(E)$ of a bundle $E$ on $S$ is defined
by $\mu_{[\omega]}(E)=\frac{1}{{\rm rank}(E)}c_1(E)\cdot_S
[\omega]$ and $[\omega]$ is an ample divisor of $S$. The
poly-stability also requires that $V_2$ is a $[\omega]$-stable
bundle. Since $S$ is a del Pezzo surface, it was shown in
\cite{BHV:2008I} that for any non-trivial holomorphic vector
bundle $E$ satisfies Eq.~(\ref{HYM00}), $h^{0}(S,E)=h^2(S,E)=0$.
This vanishing theorem dramatically simplifies the calculation of
the chiral spectrum. It turns out that the matter spectrum can be
calculated by the holomorphic Euler characteristic
\cite{Hartshone:01,Griffith:01}. By the decomposition
Eq.~(\ref{E_6 decomp}) and the vanishing theorem, the spectrum is
given by
\begin{eqnarray}
&&n_{{\bf
(3,1,1)}_{0,-4}}=-\chi(S,G^{-1})\equiv\gamma_1,\label{chi 01}
\\
&&n_{{\bf (\bar 3,1,1)}_{0,4}}=-\chi(S,G)\equiv\gamma_2,\label{chi
06}
\\
&&n_{{\bf (3,2,2)}_{0,2}}=-\chi(S,U_2)\equiv\gamma_3,\label{chi
02}\\
&&n_{{\bf (\bar
3,2,2)}_{0,-2}}=-\chi(S,U_2^{\vee})\equiv\gamma_4,\label{chi 03}
\\
&&n_{{\bf (3,2,1)}_{-3,-1}}=-\chi(S,F)\equiv\gamma_5,\label{chi
07}\\
&&n_{{\bf (\bar
3,2,1)}_{3,1}}=-\chi(S,F^{-1})\equiv\gamma_6,\label{chi 04}
\\
&&n_{{\bf (3,1,2)}_{3,-1}}=-\chi(S,U_2^{\vee}\otimes
F^{-1})\equiv\gamma_7,\label{chi 05}
\\
&&n_{{\bf (\bar 3,1,2)}_{-3,1}}=-\chi(S,U_2\otimes
F)\equiv\gamma_8,\label{chi 08}
\\
&&n_{{\bf (1,1,2)}_{-3,-3}}=-\chi(S,U_2^{\vee}\otimes
F)\equiv\delta_1,\label{chi 06}
\\
&&n_{{\bf (1,1,2)}_{3,3}}=-\chi(S,U_2\otimes
F^{-1})\equiv\delta_2,\label{chi 06}
\\
&&n_{{\bf (1,2,1)}_{-3,3}}=-\chi(S,G\otimes
F)\equiv\delta_3,\label{chi 08}
\\
&&n_{{\bf (1,2,1)}_{3,-3}}=-\chi(S,G^{-1}\otimes
F^{-1})\equiv\delta_4,\label{chi 08}
\end{eqnarray}
where $\vee$ stands for the dual bundle, $\chi$ is the holomorphic
Euler characteristic defined by $\chi(S,E)=\sum_ih^{0,i}(S,E)$,
$U_2=V_2\otimes L_2^2$, $F={L}_1^{-3}\otimes {L}_2^{-1}$,
$G={L}_{2}^{4}$, and $\gamma_{i},\;\delta_i\in
\mathbb{Z}_{\geqslant 0}$. After some algebra, Eqs.~(\ref{chi
01})-(\ref{chi 08}) can be recast as
\begin{eqnarray}
&&c_1(G)^{2}=-2-\gamma_1-\gamma_2,\label{Spetrum bundle
constraints01}\\
&&c_1(F)^{2}=-2-\gamma_5-\gamma_6,\\
&&c_1(S)\cdot c_1(G)=\gamma_1-\gamma_2,\\
&&c_1(S)\cdot c_1(F)=\gamma_6-\gamma_5,\\
&&c_2(V_2)=\frac{1}{4}(6-\gamma_1-\gamma_2+2\gamma_3+2\gamma_4),\\
&&c_1(G)\cdot
c_1(F)=\frac{1}{2}(4+\gamma_3+\gamma_4+2\gamma_5+2\gamma_6-\gamma_7-\gamma_8),\label{Spetrum
bundle constraints06}\\
&&\gamma_1-\gamma_2+\gamma_3-\gamma_4=0,\label{Spectrum constraint
01}\\
&&\gamma_1-\gamma_2-2\gamma_5+2\gamma_6-\gamma_7+\gamma_8=0,\label{Spectrum
constraint 02}\\
&&\delta_1=\frac{1}{2}(8+\gamma_1-\gamma_2+2\gamma_3+2\gamma_4+6\gamma_5+2\gamma_6-\gamma_7-\gamma_8),\label{Spectrum
delta 01}\\
&&\delta_2=\frac{1}{2}(8-\gamma_1+\gamma_2+2\gamma_3+2\gamma_4+2\gamma_5+6\gamma_6-\gamma_7-\gamma_8),\\
&&\delta_3=-\frac{1}{2}(2-2\gamma_2+\gamma_3+\gamma_4+2\gamma_6-\gamma_7-\gamma_8),\\
&&\delta_4=-\frac{1}{2}(2-2\gamma_1+\gamma_3+\gamma_4+2\gamma_5-\gamma_7-\gamma_8).\label{Spectrum
delta 04}
\end{eqnarray}
Note that given $\gamma_k,\;k=1,2,...,8$ satisfying the
constraints Eqs.~(\ref{Spectrum constraint 01}) and (\ref{Spectrum
constraint 02}), $(F,G,V_2)$ are constrained by Eqs.~(\ref{Spetrum
bundle constraints01})-(\ref{Spetrum bundle constraints06}) and
$(\delta_1,\delta_2,\delta_3,\delta_4)$ are then given by
Eqs.~(\ref{Spectrum delta 01})-(\ref{Spectrum delta 04}). In
particular, we are interested in the configurations of the
vector-like pairs, namely
$(\gamma_1,\gamma_2,\gamma_3,\gamma_4,\gamma_5,\gamma_6,\gamma_7,\gamma_8,\delta_1,\delta_2,\delta_3,\delta_4)=(a,a,b,b,c,c,d,d,e,e,f,f)$,
where $a$, $b$, $c$, $d$, $e$ are all non-negative integers. Then
Eqs.~(\ref{Spetrum bundle constraints01})-(\ref{Spectrum delta
04}) reduce to
\begin{equation}
\left\{\begin{array}{l} c_{1}(G)^{2}=-2-2a\\
c_1(F)^{2}=-2-2c
\\c_1(S)\cdot c_1(G)=0\\
c_1(S)\cdot c_1(F)=0\\
c_2(V_2)=\frac{1}{2}(3+2b-a)\\
c_1(G)\cdot c_1(F)=2+b+2c-d\\
e=4+2b+4c-d \\
f=-1+a-b-c+d.\label{Exotic free condition E_6 02}
\end{array}   \right.
\end{equation}
It was proven in \cite{Mathbundle} that for an algebraic surface
$S$ with a given $n\geqslant 4([h^0(S,K_S)/2]+1)$, there exists a
$[\omega]$-stable bundle $V$ of rank two with $c_1(V)=0$ and
$c_2(V)=n$. When $S$ is a del Pezzo surface, $h^0(S,K_S)=0$ and
this theorem implies that for any given number $m\geqslant 4$,
there exists a $[\omega]$-stable bundle of rank two with
$c_1(V)=0$ and $c_2(V)=m$. To apply this theorem to our case, we
require that $c_2(V_2)\geqslant 4$. In general, $c_1(V)$ and
$c_2(V)$ of a stable bundle $V$ over a compact K\"ahler surface
$S$ with $c_1(S)>0$ satisfy the inequality
$2rc_2(V)-(r-1)c_1(V)^2\geqslant (r^2-1)$, where $r$ is the rank
of $V$\cite{bundleinequality:01}. When $r=2$ and $c_1(V)=0$, one
can obtain the lower bound $c_2(V)\geqslant 2$. It is possible to
obtain a $[\omega]$-stable bundle $V$ of rank two with $c_1(V)=0$
and $c_2(V)\leqslant 4$ for $S$ being a del Pezzo surface. One can
start with $V$ defined by the following extension:
\begin{equation}
0\ra L\ra V\ra M\ra 0.\label{Extension}
\end{equation}
To obtain vanishing $c_1(V)$, one can set $M=L^{-1}$ and compute
$c_2(V)=-c_1(L)^2$. The extension is classified by ${\rm
Ext}^1(L,M)=H^1(S,L\otimes M^{\ast})$. When $M=L^{-1}$, the
obstruction of the non-trivial extension is $h^{1}(S,L^2)\neq 0$.
Let $L$ be a non-trivial line bundle and $S$ be a del Pezzo
surface. By the vanishing theorem, one can obtain
\begin{equation}
h^{1}(S,L^2)=-1-c_1(S)\cdot c_1(L)-2c_1(L)^2.
\end{equation}
If $c_1(S)\cdot c_1(L)=0$ with negative $c_1(L)^2$, it is easy to
see that $h^1(S,L^2)\geqslant 1$. The simple example for such a
line bundle is $L=\mathcal{O}_S(e_i-e_j),\;i\neq j$, where
$\{e_1,...,e_8\}$ is a set of the exceptional divisors of $S$.
With non-trivial extensions, one may construct a $[\omega]$-stable
bundle $V$ with $(r,c_1(V),c_2(V))=(2,0,2)$ and with the structure
group $SU(2)$ \cite{Chung:2010xx}. In what follows, we shall focus
on the case of $c_2(V_2)\geqslant 4$. We summarize the constraints
for $(a,b,c,d)$ as follows:
\begin{equation}
\left\{\begin{array}{l} 2b+4c-d \geqslant -4\\
a-b-c+d\geqslant 1\\
a-2b\leqslant -5 \\
a,b,c,d\in\mathbb{Z}_{\geqslant 0}.\label{Exotic free
conditionSO(10)}
\end{array}   \right.
\end{equation}
Note that $a$ must be odd otherwise $c_2(V_2)$ cannot be integral.
It follows from the condition $c_2(V_2)\geqslant 4$ that
$b\geqslant 3$. Let us consider the case $(a,b,c)=(1,3,0)$. Then
Eq.~(\ref{Exotic free condition E_6 02}) becomes
\begin{equation}
\left\{\begin{array}{l} c_{1}(G)^{2}=-4\\
c_1(F)^{2}=-2
\\c_1(S)\cdot c_1(G)=0\\
c_1(S)\cdot c_1(F)=0\\
c_2(V_2)=4\\
c_1(G)\cdot c_1(F)=5-d\\
e=10-d\\
f=-3+d.\label{Exotic free condition02}
\end{array}   \right.
\end{equation}
Note that for the case $(a,b,c)=(1,3,0)$, the necessary condition
for $d$ is $3\leqslant d\leqslant 10$. From the conditions
$c_1(G)^2=-4$ and $c_1(F)^2=-2$, we set
$G=\mathcal{O}_S(e_i-e_j+e_k-e_l),\;i\neq j\neq k\neq l$ and
$F=\mathcal{O}_S(e_m-e_n),\;m\neq n$. Clearly, $G$ and $F$ also
satisfy the conditions $c_1(S)\cdot c_1(G)=0$ and $c_1(S)\cdot
c_1(F)=0$. We shall not attempt to explore all solutions $(G,F)$
and only list some solutions as follows \cite{Chung:2010xx}:
\begin{equation}
(G,F)=\left\{\begin{array}{l} (\mathcal{O}_S(e_i-e_j+e_k-e_l),\mathcal{O}_S(e_i-e_j)),\;(d,e,f)=(7,3,4)\\
(\mathcal{O}_S(e_i-e_j+e_k-e_l),\mathcal{O}_S(e_m-e_j)),\;(d,e,f)=(6,4,3)
\\(\mathcal{O}_S(e_i-e_j+e_k-e_l),\mathcal{O}_S(e_i-e_k)),\;(d,e,f)=(5,5,2)\\
(\mathcal{O}_S(e_i-e_j+e_k-e_l),\mathcal{O}_S(e_j-e_n)),\;(d,e,f)=(4,6,1)\\
(\mathcal{O}_S(e_i-e_j+e_k-e_l),\mathcal{O}_S(e_j-e_k)),\;(d,e,f)=(3,7,0).\label{Exotic
free conditionSO(10)130}
\end{array}   \right.
\end{equation}

Let us consider another example, $(a,b,c)=(3,4,0)$. In this case
Eq.~(\ref{Exotic free condition E_6 02}) reduces to
\begin{equation}
\left\{\begin{array}{l} c_{1}(G)^{2}=-8\\
c_1(F)^{2}=-2
\\c_1(S)\cdot c_1(G)=0\\
c_1(S)\cdot c_1(F)=0\\
c_2(V_2)=4\\
c_1(G)\cdot c_1(F)=6-d\\
e=12-d\\
f=-2+d.\label{Exotic free condition04}
\end{array}   \right.
\end{equation}
When $(a,b,c)=(3,4,0)$, it follows from Eq.~(\ref{Exotic free
condition04}) that the necessary condition for $d$ is $2\leqslant
d\leqslant 12$. From the conditions $c_1(G)^2=-8$ and
$c_1(F)^2=-2$, we set $G=\mathcal{O}_S(2e_i-2e_j),\;i\neq j$ and
$F=\mathcal{O}_S(e_m-e_n),\;m\neq n$. It is not difficult to see
that $G$ and $F$ satisfy the conditions $c_1(S)\cdot c_1(G)=0$ and
$c_1(S)\cdot c_1(F)=0$. Some solutions of $(G,F)$ are as follows:
\begin{equation}
(G,F)=\left\{\begin{array}{l} (\mathcal{O}_S(2e_i-2e_j),\mathcal{O}_S(e_i-e_j)),\;(d,e,f)=(10,2,8)\\
(\mathcal{O}_S(2e_i-2e_j),\mathcal{O}_S(e_m-e_j)),\;(d,e,f)=(8,4,6)
\\(\mathcal{O}_S(2e_i-2e_j),\mathcal{O}_S(e_m-e_n)),\;(d,e,f)=(6,6,4)\\
(\mathcal{O}_S(2e_i-2e_j),\mathcal{O}_S(e_m-e_i)),\;(d,e,f)=(4,8,2)\\
(\mathcal{O}_S(2e_i-2e_j),\mathcal{O}_S(e_j-e_i)),\;(d,e,f)=(2,10,0).\label{Exotic
free conditionSO(10)}
\end{array}   \right.
\end{equation}

Let us turn to the chiral spectrum on the matter curves. The
breaking pattern of the presentation $\bf 27$ is
\begin{equation}
\begin{array}{ccl}
E_6 & \xrightarrow & SU(3)\times SU(2)_1\times [SU(2)_2\times
U(1)_a\times U(1)_b] \\ \\
{\bf 27} & \rightarrow &  {\bf (3,2,1)}_{1,-1}+ {\bf
(1,2,1)}_{1,3} +{\bf (\bar 3,1,2)}_{1,1}+ {\bf (1,1,2)}_{1,-3} \\
&& + {\bf (3,1,1)}_{-2,2}+{\bf (\bar 3,1,1)}_{-2,-2} +{\bf
(1,2,2)}_{-2,0} + {\bf (1,1,1)}_{4,0}.
\end{array}
\end{equation}
Let us define $V_{{\bf 27}}\otimes L_1^4|_{\Sigma^{(k)}_{\bf
27}}=\Gamma|_{\Sigma^{(k)}_{\bf 27}}=M^{(k)}$,
$F|_{\Sigma^{(k)}_{\bf 27}}=N_1^{(k)}$, and $G|_{\Sigma^{(k)}_{\bf
27}}=N_2^{(k)}$. The chirality of matter localized on matter
curves $\Sigma_{{\bf 27}}^{(k)}$ is determined by the restrictions
of the cover flux $\Gamma$ and gauge fluxes to the curves. The
spectrum induced by the cover flux and gauge fluxes is summarized
in Table \ref{Total chirality}.
\begin{table}[h]
\begin{center}
\renewcommand{\arraystretch}{.9}
\begin{tabular}{|c|c|c|c|} \hline
Curve & Matter & Bundle & Chirality  \\ \hline
\multirow{8}{*}{${\bf 27}^{(k)}$} & ${\bf (3,2,1)}_{1,-1}$ &
$V_{{\bf 27}}\otimes L_1\otimes
L_2^{-1}|_{\Sigma^{(k)}_{\bf 27}}$ & $M^{(k)}+N_1^{(k)}$\\
& ${\bf (1,2,1)}_{1,3}$ & $V_{{\bf 27}}\otimes L_1\otimes
 L_2^{3}|_{\Sigma^{(k)}_{\bf 27}}$ & $M^{(k)}+N_1^{(k)}+N_2^{(k)}$ \\
& ${\bf (\bar 3,1,2)}_{1,1}$ & $V_{{\bf 27}}\otimes V_2\otimes
L_1\otimes
 L_2|_{\Sigma^{(k)}_{\bf 27}}$ & $2(M^{(k)}+N_1^{(k)})+N_2^{(k)}$ \\
& ${\bf (1,1,2)}_{1,-3}$ & $V_{{\bf 27}}\otimes V_2\otimes
L_1\otimes
 L_2^{-3}|_{\Sigma^{(k)}_{\bf 27}}$ & $2(M^{(k)}+N_1^{(k)})-N_2^{(k)}$\\
& ${\bf (3,1,1)}_{-2,2}$ & $V_{{\bf 27}}\otimes L_1^{-2}\otimes
 L_2^{2}|_{\Sigma^{(k)}_{\bf 27}}$ & $M^{(k)}+2N_1^{(k)}+N_2^{(k)}$ \\
& ${\bf (\bar 3,1,1)}_{-2,-2}$ & $V_{{\bf 27}}\otimes
L_1^{-2}\otimes
 L_2^{-2}|_{\Sigma^{(k)}_{\bf 27}}$ & $M^{(k)}+2N_1^{(k)}$ \\
& ${\bf (1,2,2)}_{-2,0}$ & $V_{{\bf 27}}\otimes V_2\otimes
L_1^{-2}|_{\Sigma^{(k)}_{\bf 27}}$ & $2(M^{(k)}+2N_1^{(k)})+N_2^{(k)}$ \\
& ${\bf (1,1,1)}_{4,0}$ & $V_{{\bf 27}}\otimes
L_1^{4}|_{\Sigma^{(k)}_{\bf 27}}$ & $M^{(k)}$  \\\hline
\end{tabular}
\caption{Chirality of matter localized on matter curve ${\bf
27}^{(k)}$.} \label{Total chirality}
\end{center}
\end{table}

\section{Tadpole Cancellation}

The cancellation of tadpoles is crucial for consistent
compactifications. In general, there are induced tadpoles from
7-brane, 5-brane, and 3-brane charges in F-theory. The 7-brane
tadpole cancellation in F-theory is automatically satisfied since
$X_4$ is a Calabi-Yau manifold. The cancellation of the $D5$-brane
tadpole in the spectral cover construction follows from the
topological condition that the overall first Chern class of the
Higgs bundle vanishes. Therefore, the non-trivial tadpole
cancellation in F-theory needed to be satisfied is the $D3$-brane
tadpole which can be calculated by the Euler characteristic
$\chi(X_4)$. The cancellation condition is of the form
\cite{Sethi:1996es}
\begin{equation}
N_{D3}=\frac{{\chi}(X_4)}{24}-\frac{1}{2}\int_{X_4}G\wedge
G,\label{Tadpole cancellation}
\end{equation}
where $N_{D3}$ is the number of $D3$-branes and $G$ is the
four-form flux on $X_4$. For a non-singular elliptically fibered
Calabi-Yau fourfold $X_4$, it was shown in \cite{Sethi:1996es}
that the Euler characteristic $\chi(X_4)$ can be expressed as
\begin{equation}
\chi(X_4)=12\int_{B_3}c_1(B_3)[c_2(B_3)+30c_1(B_3)^2],\label{Euler
number smooth}
\end{equation}
where $c_k(B_3)$ are the Chern classes of $B_3$. It follows from
Eq.~(\ref{Euler number smooth}) that $\chi(X_4)/24$ is at least
half-integral\footnote{For a generic Calabi-Yau manifold $X_4$,
$\chi(X_4)/24$ takes value in $\mathbb{\mathbb{Z}}_4$
\cite{Sethi:1996es}.}. When $X_4$ admits non-abelian
singularities, the Euler characteristic of $X_4$ is replaced by a
refined Euler characteristic, the Euler characteristic of the
smooth fourfold obtained from a suitable resolution of $X_4$. On
the other hand, $G$-flux encodes the two-form gauge fluxes on the
7-branes. It was shown in \cite{Curio:1998bva} that
\begin{equation}
\int_{X_4} G\wedge G=-\Gamma^{2},
\end{equation}
where $\Gamma$ is the universal cover flux defined in section 2
and $\Gamma^2$ is defined as the self-intersection number of
$\Gamma$ inside the spectral cover. It is a challenge to find
compactifications with non-vanishing $G$-flux and non-negative
$N_{D_3}$ to satisfy the tadpole cancellation condition
(\ref{Tadpole cancellation}). In the next two subsections, we
shall derive the formulae of the refined Euler characteristic
$\chi(X_4)$ and the self-intersection of the universal cover
fluxes $\Gamma^2$ for the $(2,1)$ and $(1,1,1)$ factorizations.

\subsection{Geometric Contribution}

In the presence of non-abelian singularities, $X_4$ becomes
singular and the Euler characteristic $\chi(X_4)$ needs to be
modified by resolving the singularities. To be more concrete, let
us define $H$ to be the gauge group corresponding to the
non-abelian singularity over $S$ and $G$ to be the complement of
$H$ in $E_8$. Then the Euler characteristic is modified to
\begin{equation}
{\chi}(X_4)=\chi^{\ast}(X_4)+\chi_G-\chi_{E_8},\label{refined chi}
\end{equation}
where $\chi^{\ast}(X_4)$ is the Euler characteristic for a smooth
fibration over $B_3$ given by Eq.~(\ref{Euler number smooth}) and
the characteristic $\chi_{E_8}$ is given by \cite{Andreas:1999ng,
Curio:1998bva, Blumenhagen:global02}
\begin{equation}
\chi_{E_8}=120\int_S(3\eta^2-27\eta c_1+62c_1^2).
\end{equation}
For the case of $G=SU(n)$, the characteristic $\chi_{SU(n)}$ is
computed as
\begin{equation}
\chi_{SU(n)}=\int_S(n^3-n)c_1^2+3n\eta(\eta-nc_1).
\end{equation}
When the group $G$ splits into a product of two groups $G_1$ and
$G_1$, $\chi_{G}$ in Eq.~(\ref{refined chi}) is then replaced by
$\chi^{(k)}_{G_1}+\chi^{(l)}_{G_2}$ where $\eta$ in $\chi_{G}$ is
split into the classes $\eta^{(m)}$ as shown in the footnote
below. It turns out that the refined Euler characteristic of the
$(2,1)$ factorization is given by
\begin{eqnarray}
{\chi}(X_4)&=&\chi^{\ast}(X_4)+\chi^{(a)}_{SU(2)}+\chi^{(b)}_{SU(1)}
-\chi_{E_8}\nonumber\\&=&\chi^{\ast}(X_4)+\int_S3[c_1(32c_1-16t-15\xi)+(2t^2+4t\xi+3\xi^2)]-\chi_{E_8}.\label{refined
chi2,1}
\end{eqnarray}
In the $(1,1,1)$ factorization, the refined Euler
characteristic\footnote{For the $(2,1)$ factorization,
$\eta^{(a)}=(\eta-c_1-\xi)$ and $\eta^{(b)}=(c_1+\xi)$. For the
$(1,1,1)$ factorization, $\eta^{(l_1)}=(\eta-2c_1-\xi_1-\xi_2)$,
$\eta^{(l_2)}=(c_1+\xi_1)$, and $\eta^{(l_3)}=(c_1+\xi_2)$.} is
\begin{eqnarray}
{\chi}(X_4)&=&\chi^{\ast}(X_4)+\chi^{(l_1)}_{SU(1)}+\chi^{(l_2)}_{SU(1)}+\chi^{(l_3)}_{SU(1)}-\chi_{E_8}\nonumber\\&=&
\chi^{\ast}(X_4)+\int_S3\{c_1[12c_1-7t-6(\xi_1+\xi_2)]+[t^2+2t(\xi_1+\xi_2)+2(\xi_1^2+\xi_1\xi_2+\xi_2^2)]\}\nonumber\\&-&\chi_{E_8}.\label{refined
chi1,1,1}
\end{eqnarray}

\subsection{Cover Flux Contribution}

Under cover factorizations, the universal cover flux is of the
form
\begin{equation}
\Gamma=\sum_{k}\Gamma^{(k)},
\end{equation}
where the fluxes $\Gamma^{(k)}$ satisfy the traceless condition
$\sum_{k}p_{k\ast}\Gamma^{(k)}=0$. In what follows, we shall
compute the self-intersection $\Gamma^2$ of the universal fluxes
for the (2,1) and (1,1,1) factorizations.

\subsubsection{$(2,1)$ Factorization}

Let us recall that in the $(2,1)$ factorization, the universal
cover flux is given by
\begin{equation}
\Gamma=k_a \gamma^{(a)}_0+k_b\gamma^{(b)}_0+m_a \delta^{(a)} +m_b
\delta^{(b)} + \tilde{\rho}=\Gamma^{(a)}+\Gamma^{(b)},
\end{equation}
where $\Gamma^{(a)}$ and $\Gamma^{(b)}$ are
\begin{equation}
\Gamma^{(a)}=[\mathcal{C}^{(a)}]\cdot \left[(2k_a+m_a)\sigma
-\pi^{\ast}(k_a [a_2] +m_b[d_1]+\rho)\right]
\equiv[\mathcal{C}^{(a)}]\cdot[\widetilde{\mathcal{C}}^{(a)}],\label{dual
cover a}
\end{equation}
\begin{equation}
\Gamma^{(b)}=[\mathcal{C}^{(b)}]\cdot \left[(k_b+2m_b)\sigma
-\pi^{\ast}(k_b[d_1]+m_a[a_2]-2\rho)
\right]\equiv[\mathcal{C}^{(b)}]\cdot[\widetilde{\mathcal{C}}^{(b)}].\label{dual
cover b}
\end{equation}
Then the self-intersection $\Gamma^2$ is calculated by
\cite{Caltech:global03}
\begin{equation}
\Gamma^2=[\mathcal{C}^{(a)}]\cdot[\widetilde{\mathcal{C}}^{(a)}]\cdot[\widetilde{\mathcal{C}}^{(a)}]+[\mathcal{C}^{{(b)}}]\cdot[\widetilde{\mathcal{C}}^{(b)}]
\cdot[\widetilde{\mathcal{C}}^{(b)}].
\end{equation}
Recall that in the $(2,1)$ factorization,
$[\mathcal{C}^{(a)}]=2\sigma+\pi^{\ast}(\eta-c_1-\xi)$ and
$[\mathcal{C}^{(b)}]=\sigma+\pi^{\ast}(c_1+\xi)$. With
Eqs.~(\ref{dual cover a}) and (\ref{dual cover b}), it is
straightforward to compute
\begin{eqnarray}
&\Gamma^2&=[\mathcal{C}^{(a)}_2]\cdot[\tilde{\mathcal{C}}^{(a)}_2]^2
+[\mathcal{C}^{(b)}_1]\cdot[\tilde{\mathcal{C}}^{(b)}_1]^2
\nonumber \\ && = -\frac{1}{2}(2k_a+m_a)^2[a_2]\cdot[a_0]
-(k_b+2m_b)^2[d_1]\cdot[d_0] \nonumber \\ &&+
\frac{3}{2}(m_a[a_2]-2m_b[d_1]-2\rho)^2.
\end{eqnarray}

\subsubsection{ $(1,1,1)$ Factorization}

In the $(1,1,1)$ factorization, the universal flux is given by
\begin{equation}
\Gamma=k_{l_1} \gamma_0^{(l_1)} +k_{l_2} \gamma_0^{(l_2)}+k_{l_3}
\gamma_0^{(l_3)} + m_{l_1} \delta^{(l_1)}+ m_{l_2} \delta^{(l_2)}+
m_{l_3} \delta^{(l_3)}
+\tilde{\rho}\equiv\Gamma^{(l_1)}+\Gamma^{(l_2)}+\Gamma^{(l_3)},
\end{equation}
where $\Gamma^{(l_1)}$, $\Gamma^{(l_2)}$, and $\Gamma^{(l_3)}$ are
\begin{eqnarray}
\Gamma^{(l_1)} \equiv [\mathcal{C}^{(l_1)}]\cdot
[\tilde{\mathcal{C}}^{(l_1)}] =[\mathcal{C}^{(l_1)}]\cdot [
(k_{l_1}+2m_{l_1})\sigma
-\pi^{\ast}( k_{l_1}[f_1] +m_{l_2}\xi_1 +m_{l_3}\xi_2 +\rho_1-\rho_3 ) ]\label{dual l1},~~ \\
\Gamma^{(l_2)} \equiv [\mathcal{C}^{(l_2)}]\cdot
[\tilde{\mathcal{C}}^{(l_2)}] = [\mathcal{C}^{(l_2)}]\cdot [
(k_{l_2}+2m_{l_2})\sigma -\pi^{\ast}( m_{l_1}[f_1] +k_{l_2}\xi_1
+m_{l_3}\xi_2 +\rho_2-\rho_1 )
],\label{dual l2}~~ \\
\Gamma^{(l_3)} \equiv [\mathcal{C}^{(l_3)}]\cdot
[\tilde{\mathcal{C}}^{(l_3)}] = [\mathcal{C}^{(l_3)}]\cdot [
(k_{l_3}+2m_{l_3})\sigma -\pi^{\ast}( m_{l_1}[f_1] +m_{l_2}\xi_1
+k_{l_3}\xi_2 +\rho_3-\rho_2 ) ]\label{dual l3}.~~
\end{eqnarray}
In this case the self-intersection $\Gamma^2$ is computed as
\begin{equation}
\Gamma^2=
[\mathcal{C}^{{(l_1)}}]\cdot[\widetilde{\mathcal{C}}^{(l_1)}]\cdot[\widetilde{\mathcal{C}}^{(l_1)}]+[\mathcal{C}^{(l_2)}]\cdot[\widetilde{\mathcal{C}}^{{(l_2)}}]
\cdot[\widetilde{\mathcal{C}}^{(l_2)}]+[\mathcal{C}^{(l_2)}]\cdot[\widetilde{\mathcal{C}}^{{(l_3)}}]
\cdot[\widetilde{\mathcal{C}}^{(l_3)}].
\end{equation}
Recall that
$[\mathcal{C}^{(l_1)}]=\sigma+\pi^{\ast}(\eta-2c_1-\xi_1-\xi_2)$,
$[\mathcal{C}^{(l_2)}]=\sigma+\pi^{\ast}(c_1+\xi_1)$, and
$[\mathcal{C}^{(l_3)}]=\sigma+\pi^{\ast}(c_1+\xi_2)$. It follows
from Eqs.~(\ref{dual l1})-(\ref{dual l3}) that
\begin{eqnarray}
\Gamma^2&=&[\mathcal{C}^{(l_1)}]\cdot[\tilde{\mathcal{C}}^{(l_2)}]^2
+[\mathcal{C}^{(l_2)}]\cdot[\tilde{\mathcal{C}}^{(l_2)}]^2
+[\mathcal{C}^{(l_3)}]\cdot[\tilde{\mathcal{C}}^{(l_3)}]^2
\nonumber \\ & =& -(k_{l_1}+2m_{l_1})^2[f_1]\cdot[f_0]
-(k_{l_2}+2m_{l_2})^2[g_1]\cdot[g_0]
-(k_{l_3}+2m_{l_3})^2[h_1]\cdot[h_0] \nonumber \\ &&+
(\rho_1-\rho_3-2m_{l_1}[f_1]+m_{l_2}[g_1]+m_{l_3}[h_1])^2 \nonumber \\
&&+ (\rho_2-\rho_1+m_{l_1}[f_1]-2m_{l_2}[g_1]+m_{l_3}[h_1])^2 \nonumber \\
&&+ (\rho_3-\rho_2+m_{l_1}[f_1]+m_{l_2}[g_1]-2m_{l_3}[h_1])^2.
\end{eqnarray}


\section{Models}

In this section we give some numerical examples in the geometric
backgrounds $dP_2$ studied in \cite{Caltech:global01} and $dP_7$
in \cite{Blumenhagen:global02}. The basic geometric data of $dP_2$
in $X_4$ is
\begin{eqnarray}
c_1=3h-e_1-e_2, ~~ t=h, ~~ \eta=17h-6e_1-6e_2.
\end{eqnarray}
It follows from Eqs.~(\ref{refined chi2,1}) and (\ref{refined
chi1,1,1}) that the refined Euler characteristic $\chi(X_4)$ for
the $(2,1)$ and $(1,1,1)$ factorizations are
\begin{eqnarray}
\chi(X_4)_{(2,1)}&=&10662+\int_S3[-15\xi c_1+4t\xi+3\xi^2],\\
\chi(X_4)_{(1,1,1)}&=&10320+\int_S6\left[(t-3c_1)(\xi_1+\xi_2)+(\xi_1^2+\xi_1\xi_2+\xi_2^2)\right],
\end{eqnarray}
where $\chi^{\ast}(X_4)=13968$ has been used. The ample divisor
$[\omega]_{dP_2}$ is chosen to be
\begin{equation}
[\omega]_{dP_2}=\alpha(e_1+e_2)+\beta(h-e_1-e_2),
~~~2\alpha>\beta>\alpha>0.
\end{equation}
For the $dP_7$ studied in \cite{Blumenhagen:global02}, the basic
geometric data is
\begin{eqnarray}
c_1&=&3h-e_1-e_2-e_3-e_4-e_5-e_6-e_7,     \nonumber\\
t&=&2h-e_1-e_2-e_3-e_4-e_5-e_6, \\
\eta&=&16h-5e_1-5e_2-5e_3-5e_4-5e_5-5e_6-6e_7. \nonumber
\end{eqnarray}
with $\chi^{\ast}(X_4)=1728$. By Eqs.~(\ref{refined chi2,1}) and
(\ref{refined chi1,1,1}), the refined Euler characteristic
$\chi(X_4)$ for the $(2,1)$ and $(1,1,1)$ factorizations are
\begin{eqnarray}
\chi(X_4)_{(2,1)}&=&708+\int_S3[-15\xi c_1+4t\xi+3\xi^2],\\
\chi(X_4)_{(1,1,1)}&=&594+\int_S6\left[(t-3c_1)(\xi_1+\xi_2)+(\xi_1^2+\xi_1\xi_2+\xi_2^2)\right].
\end{eqnarray}
In this case we choose the ample divisor $[\omega]_{dP_7}$ to be
\begin{equation}
[\omega]_{dP_7}=14\beta h -(5\beta-\alpha)\sum_{i=1}^7 e_i,
~~~5\beta>\alpha>0.
\end{equation}
We shall discuss the models of the (2,1) and (1,1,1)
factorizations. In each case the trivial and non-trivial
restrictions of the $U(1)$ fluxes to the matter curves will be
discussed. Non-trivial restriction leads to the modification of
the chirality of each matter on the curve after $E_6$ is broken
according to the calculation in section 3. In addition, there
could exist vector-like pairs on each curve since we only know the
net chirality. The Higgs vector-like pair $({\bf 27}+\overline{\bf
27})$ needed for the gauge unification is therefore assigned to
one of these pairs, though the machinery to calculate the exact
number of these vector-like fields is not clear yet.

\subsection{Examples of the $(2,1)$ Factorization}

In the (2,1) factorization the matter fields are assigned to ${\bf
27}^{(a)}$ curve and the Higgs fields come from the other ${\bf
27}^{(b)}$ curve. The Yukawa coupling then turns out to be
\begin{equation}
\mathcal{W}\supset {\bf 27}^{(a)} \cdot{\bf 27}^{(a)} \cdot {\bf
27}^{(b)}.
\end{equation}
Since the fermion and Higgs fields are not on the same $\bf 27$
curve, the exotic fields in ${\bf 27}^{(a)}$ can be taken as
exotic quarks and leptons which are able to mix with the ordinary
ones by suitable discrete symmetries and to decay via mechanisms
such as FCNC after $E_6$ is broken mentioned in section 3.

\subsubsection{A three-family $E_6$ model in $dP_2$}

The parameters of the model are listed in Table \ref{table
Model01}.
\begin{table}[h]
\center
\renewcommand{\arraystretch}{.9}
\begin{tabular}{|c|c|c|c|c||c||c|c|}
\hline $k_a$ & $k_b$ & $m_a$ & $m_b$ & $\rho$ & $\xi$ & $\alpha$ &
$\beta$ \\ \hline

0.5 & -1.5 & -1 & -1 & $-\frac{5}{2}h+\frac{3}{2}e_1-\frac{3}{2}e_2$ & $e_1$ & 2 & 3 \\
\hline
\end{tabular}
\caption{Parameters of an example of a three-generation $E_6$ GUT.}\label{table Model01} 
\end{table}

These parameters give the spectrum $N_{{\bf 27}^{(a)}}=3$ and
$N_{{\bf 27}^{(b)}}=3$ with $N_{D3}=415$ as shown in Table
\ref{T-21dP2-3}. The $dP_2$ surface is probably too limited for
the fluxes to break the $E_6$ gauge group. Therefore, we stop at a
three-generation $E_6$ GUT model in this example.

\begin{table}[h]
\renewcommand{\arraystretch}{.9}
\center
\begin{tabular}{|c|c|c|}
\hline Curve & Class & Gen.\\ \hline

${\bf 27}^{(a)}$ & $8h-4e_1-3e_2$ & 3 \\ \hline

${\bf 27}^{(b)}$ & $e_1$ & 3 \\ \hline
\end{tabular}
\caption{The $\bf 27$ curves of the three-generation $E_6$ example in $dP_2$.} \label{T-21dP2-3} 
\end{table}

\subsubsection{An example of three-generation without flux restriction in $dP_7$}

The parameters of the model with $N_{D3}=12$ are listed in the
Table \ref{21-dP7noflux}.
\begin{table}[h]
\center
\renewcommand{\arraystretch}{.9}
\begin{tabular}{|c|c|c|c|c||c||c|c|}
\hline $k_a$ & $k_b$ & $m_a$ & $m_b$ & $\rho$ & $\xi$ & $\alpha$ &
$\beta$ \\ \hline

-0.5 & 1.5 & 0 & -0.5 & $\frac{1}{2}(3e_1+e_2+e_3+e_4)$ & $h-e_5-e_6+e_7$ & 3 & 1 \\
\hline
\end{tabular}
\caption{Parameters of an example of the (2,1) factorization in $dP_7$.} \label{21-dP7noflux} 
\end{table}

The matter contents on the curves are listed in Table
\ref{T-21dP7-3}. If the line bundles $G$ and $F$ associated to
$SU(2)\times U(1)_a \times U(1)_b$ flux are chosen to have trivial
restrictions\footnote{To avoid receiving a Green-Schwarz mass, it
is required that $[H]\cdot_S c_1$=0 and $[H]\cdot_S\eta=0$, for
$H=F,\;G$
\cite{BHV:2008II,Donagi:2008ll,Donagi:2008sl,BHV:2008I,Caltech:global02}.}
to both matter {\bf 27} curves, for example,
$F=\mathcal{O}_S(e_5-e_6)$ and
$G=\mathcal{O}_S(e_1-e_2+e_3-e_4)$,\footnote{$G$ can be chosen
also as $G=\mathcal{O}_S(2(e_3-e_4))$ from Eq.~(\ref{Exotic free
conditionSO(10)}).} then the chirality on each matter curve
remains the same after $E_6$ is broken down to $SU(3)\times
SU(2)\times U(1)_a\times U(1)_b$. After suitably transforming the
$U(1)$ gauge groups, the corresponding matter content and
phenomenology at low energy is a conventional rank 5 model
discussed in section 3.
\begin{table}[h]
\renewcommand{\arraystretch}{.9}
\center
\begin{tabular}{|c|c|c|}
\hline Curve & Class & Gen.\\ \hline

${\bf 27}^{(a)}$ & $6h-2e_1-2e_2-2e_3-2e_4-e_5-e_6-4e_7$ & 3 \\
\hline

${\bf 27}^{(b)}$ & $h-e_5-e_6+e_7$ & 2 \\ \hline
\end{tabular}
\caption{The $\bf 27$ curves of the example of the (2,1)
factorization without flux restrictions
in $dP_7$.} \label{T-21dP7-3} 
\end{table}

\subsubsection{An example with non-trivial flux restrictions in $dP_7$}

In this example we consider a model with non-trivial flux
restrictions to the matter curves in $dP_7$. From the chirality
formulae discussed in section 3 and listed in Table \ref{Total
chirality}, we find that it is unavoidable to have exotic fields
under this construction. To maintain at least three copies for the
MSSM matter after the gauge group $E_6$ is broken, we may have to
start from a model with more chirality on the $\bf 27$ curves. The
parameters of an example of this scenario are listed in Table
\ref{table01}.
\begin{table}[h]
\center
\renewcommand{\arraystretch}{.9}
\begin{tabular}{|c|c|c|c|c||c||c|c|}
\hline $k_a$ & $k_b$ & $m_a$ & $m_b$ & $\rho$ & $\xi$ & $\alpha$ &
$\beta$ \\ \hline

0.5 & -0.5 & -1 & -0.5 & $-h+\frac{1}{2}(e_1-2e_2+e_3+e_4+e_6)$ & $h-e_2+e_5-e_7$ & 13 & 11 \\
\hline
\end{tabular}
\caption{Parameters of an example with non-trivial flux restrictions in $dP_7$.}\label{table01} 
\end{table}

It follows from Eq.~(\ref{Tadpole cancellation}) and the
parameters in Table \ref{table01} that $N_{D3}=14$. We choose
chirality-three curve for the matter fields and a chirality-four
curve for the Higgs fields to make sure that there are enough MSSM
matter after the gauge group $E_6$ is broken. From
Eq.~(\ref{Exotic free conditionSO(10)130}), we can turn on the
fluxes
 $F=\mathcal{O}_S(e_1-e_2)$ and
$G=\mathcal{O}_S(e_2-e_3+e_4-e_5)$ in $dP_7$.\footnote{$G$ can be
chosen also as $G=\mathcal{O}_S(2(e_4-e_5))$ from Eq.~(\ref{Exotic
free conditionSO(10)}).} The detailed information of the curves
and the restrictions of fluxes to each curve are listed in Table
\ref{T-21dP7-m}.

\begin{table}[h]
\center
\renewcommand{\arraystretch}{.9}
\begin{tabular}{|c|c|c||c|c|}
\hline Curve & Class & $M$ & $N_1$ & $N_2$ \\
\hline

${\bf 27}^{(a)}$ & $6h-2e_1-e_2-2e_3-2e_4-3e_5-2e_6-2e_7$ & 3 & 1 & -2 \\
\hline

${\bf 27}^{(b)}$ & $h-e_2+e_5-e_7$ & 4 & -1 & 2 \\ \hline
\end{tabular}
\caption{The $\bf 27$ curves with non-trivial flux restrictions in
$dP_7$.} \label{T-21dP7-m}
\end{table}

The low energy spectrum is listed in Table \ref{MSSM21dP7}.  One
can see that there are exotic fields including extra generations
of quarks. One possible solution to these exotic fields is
including them in the FCNC and CC mechanisms discussed in section
3 so that they could gain large masses and decay after mixing with
ordinary generations. The detailed low energy physics is dedicated
to future study.

\begin{table}[h]
\center
\renewcommand{\arraystretch}{.9}
\begin{tabular}{|c|c|c|c|} \hline
Rep. & Gen. on ${\bf 27}^{(a)}$  & Gen. on ${\bf 27}^{(b)}$   \\
\hline

${\bf (3,2)}_{1,-1}$ & $3\times Q+1\times {\bf (3,2)}_{1,-1}$ & 3
\\ \hline

${\bf (\bar 3,1)}_{-2,-2}$ & $3\times u^c+2\times{\bf (\bar 3,1)}_{-2,-2}$ & 2 \\
\hline

${\bf (\bar 3,1)}_{1,1}$ & $3\times d^c+3\times D$ & 4+4 \\
\hline

${\bf (1,2)}_{-2,0}$ & $3\times L+ 5\times h$ & $3\times(H_1+H_2)$ \\
\hline

${\bf (1,1)}_{4,0}$ &  $3\times e^c$ & 4 \\
\hline

${\bf (1,1)}_{1,-3}$ & $3\times \nu^c+7\times S$ & $2\times(H_3+H_4)$  \\
\hline

${\bf (3,1)}_{-2,2}$ & $3\times\bar D$ &  4 \\
\hline

${\bf (1,2)}_{1,3}$ & $2\times \bar h $ & $5\times \bar H_2$ \\
\hline

\end{tabular}
\caption{The MSSM spectrum of the $(2,1)$ factorization in
$dP_7$.} \label{MSSM21dP7}
\end{table}

\subsection{Examples of the $(1,1,1)$ Factorization}

The Yukawa coupling of the ${\bf 27}$ curves in the $(1,1,1)$
factorization is ${\bf 27}^{(l_1)} {\bf 27}^{(l_2)} {\bf
27}^{(l_3)}$. The fermions are assigned on the two $\bf 27$ curves
while the Higgs fields are located on the third $\bf 27$ curve.
For instance,
\begin{equation}
\mathcal{W}\supset {\bf 27}^{(l_1)}_M \cdot{\bf 27}^{(l_2)}_M
\cdot {\bf 27}^{(l_3)}_H.
\end{equation}
In this scenario the fermions are separated on different matter
curves and the sum of the generations should accomplish a
three-family model, for example, two families on ${\bf
27}^{(l_1)}$ and one family on ${\bf 27}^{(l_2)}$, or vice versa.
However, this construction generally results in some problems in
the mass matrices. With the assistance from the flux restrictions,
the method studied in \cite{Font:2008id} can be applied to obtain
a more reasonable Yukawa structure. However, again from the
chirality given in Table \ref{Total chirality} we expect exotic
fields to remain in the spectrum after this mechanism. In what
follows, we demonstrate one example for each case in the $(1,1,1)$
factorization.

\subsubsection{An example of three-generation without flux restriction in $dP_7$}

The parameters of the model are listed in Table \ref{table
Model02}.
\begin{table}[h]
\center
\begin{tabular}{|c|c|c|c|c|c|c||c|c||c|c|}
\hline $k_{l_1}$ & $k_{l_2}$ & $k_{l_3}$ & $m_{l_1}$ & $m_{l_2}$ &
$m_{l_3}$ & $\rho_1$ & $\xi_1$ & $\xi_2$ & $\alpha$ & $\beta$ \\
\hline

-1.5 & -0.5 & 1.5 & 0 & 0 & 0 &  $-h+e_1+2e_2$ & $e_1$ & $2h-2e_1-e_2+e_3-e_7$ & 1 & 3 \\
\hline
\end{tabular}
\caption{Parameters of a three family model in $dP_7$ with
$\rho_2=2\rho_1$ and $\rho_3=0$.}\label{table Model02} 
\end{table}

\begin{table}[h]
\center
\begin{tabular}{|c|c|c|c|}
\hline Curve & Class & Gen. & Matter\\ \hline

${\bf 27}^{(l_1)}$ & $5h-e_1-e_2-3e_3-2e_4-2e_5-2e_6-2e_7$ & 2 & Fermion \\
\hline

${\bf 27}^{(l_2)}$ & $e_1$ & 1 & Fermion \\ \hline

${\bf 27}^{(l_3)}$ & $2h-2e_1-e_2+e_3-e_7$ & 4 & Higgs \\ \hline
\end{tabular}
\caption{The spectrum of the three-generation model in $dP_7$.}
\label{T-111dP7-3}
\end{table}

These parameters give the spectrum shown in Table \ref{T-111dP7-3}
with $N_{D3}=10$. Let us choose the line bundles to be
$F=\mathcal{O}_S(e_5-e_6)$ and
$G=\mathcal{O}_S(e_2-e_4+e_3-e_6)$,\footnote{$G$ can be chosen
also as $G=\mathcal{O}_S(2(e_4-e_5))$ from Eq.~(\ref{Exotic free
conditionSO(10)}).} having trivial restrictions to each ${\bf 27}$
curve. Then the chirality remains the same after $E_6$ is broken
down to $SU(3)\times SU(2)\times U(1)_a\times U(1)_b$. After
suitably transforming the $U(1)$ charges, the corresponding matter
content and phenomenology at low energy is again a conventional
rank 5 model.

\subsubsection{An Example of non-trivial flux restrictions in $dP_7$}

The parameters of the model are listed in Table \ref{table
Model03}.
\begin{table}[h]
\center
\begin{tabular}{|c|c|c|c|c|c|c||c|c||c|c|}
\hline $k_{l_1}$ & $k_{l_2}$ & $k_{l_3}$ & $m_{l_1}$ & $m_{l_2}$ &
$m_{l_3}$ & $\rho_1$ & $\xi_1$ & $\xi_2$ & $\alpha$ & $\beta$ \\
\hline

-0.5 & -0.5 & -0.5 & 0 & 0 & -1 &  $e_2$ & $2h-2e_1-e_3-e_7$ & $h-e_1-e_2$ & 1 & 3 \\
\hline
\end{tabular}
\caption{Parameters of a three family model in $dP_7$ with
$\rho_2=2\rho_1$ and $\rho_3=0$.}\label{table Model03} 
\end{table}

These parameters confine the spectrum of $E_6$ shown in Table
\ref{T-111dP7-m} with $N_{D3}=10$. If the line bundles associated
to $SU(2)\times U(1)_a\times U(1)_b$ flux are chosen as
$F=\mathcal{O}_S(e_3-e_5)$ and
$G=\mathcal{O}_S(e_1-e_2+e_4-e_6)$,\footnote{$G$ can be chosen
also as $G=\mathcal{O}_S(2(e_3-e_4))$ from Eq.~(\ref{Exotic free
conditionSO(10)}).} then the chirality of MSSM matter after $E_6$
is broken will be modified by numbers $N_1$ and $N_2$ shown in
Table \ref{T-111dP7-m}.

\begin{table}[h]
\center
\begin{tabular}{|c|c|c||c|c|c|}
\hline Curve & Class & $M$ & $N_1$ & $N_2$ & Matter\\
\hline

${\bf 27}^{(l_1)}$ & $4h+e_1-e_2-e_3-2e_4-2e_5-2e_6-2e_7$ & 3 & -1 & -2& Fermion \\
\hline

${\bf 27}^{(l^2)}$ & $2h-2e_1-e_3-e_7$ & 0 & 1 & 2 & Fermion \\
\hline

${\bf 27}^{(l_3)}$ & $h-e_1-e_2$ & 4 & 0 & 0 & Higgs \\ \hline
\end{tabular}
\caption{The spectrum of the three-generation model in $dP_7$.} \label{T-111dP7-m} 
\end{table}

Originally, there is no chirality on curve ${\bf 27}^{(l_2)}$ so
it does not look realistic before the $E_6$ gauge group is broken.
However after the fluxes are turned on, the chirality is
``reshuffled'' and shared between curves ${\bf 27}^{(l_1)}$ and
${\bf 27}^{(l_2)}$. Therefore, we can interpret the model in the
way studied in \cite{Font:2008id} that is able to give a rich
structure to the mass matrices via the Yukawa couplings. We
demonstrate the corresponding MSSM spectrum in Table
\ref{MSSM111dP7}.

\begin{table}[h]
\center
\renewcommand{\arraystretch}{.9}
\begin{tabular}{|c|c|c|c|} \hline
Rep. & Gen. on ${\bf 27}^{(l_1)}$  & Gen. on ${\bf 27}^{(l_2)}$ & Gen. on ${\bf 27}^{(l_3)}$ \\
\hline

${\bf (3,2)}_{1,-1}$ & $2\times Q$ & $1\times Q$ & 4 \\ \hline

${\bf (\bar 3,1)}_{-2,-2}$ & $1\times u^c$ & $2\times u^c$ &  4 \\
\hline

${\bf (\bar 3,1)}_{1,1}$ & $1\times d^c+1\times D$ & $2\times d^c+2\times D$ & 8 \\
\hline

${\bf (1,2)}_{-2,0}$ & $0$ & $3\times L+ 3\times h$ & $4\times (H_1+H_2)$ \\
\hline

${\bf (1,1)}_{4,0}$ & $3\times e^c$ & $0$ & 4 \\
\hline

${\bf (1,1)}_{1,-3}$ & $3\times \nu^c+3\times S$ & $0$ & $4\times(H_3+H_4)$ \\
\hline

${\bf (3,1)}_{-2,2}$ & $1\times {\bf (\bar 3,1)}_{2,-2}$ & $3\times \bar D+1\times {\bf (3,1)}_{-2,2}$ & 4 \\
\hline

${\bf (1,2)}_{1,3}$ & $0$ & $3\times \bar h$ & $4\times \bar H_2$ \\
\hline

\end{tabular}
\caption{The MSSM matter shared by two curves in $dP_7$.}
\label{MSSM111dP7}
\end{table}

\section{Conclusions}

In this paper we discuss the $E_6$ GUT models where the gauge
group is broken by the non-abelian flux $SU(2)\times U(1)^2$ in
F-theory. The non-abelian part $SU(2)$ of the flux is not
commutative with $E_6$ so the gauge group after breaking is
$SU(3)\times SU(2)_L\times U(1)_a\times U(1)_b$ which is
equivalent to a rank-5 model with $SU(3)\times SU(2)_L\times
U(1)_Y\times U(1)_{\eta}$. We start building models from the
$SU(3)$ spectral cover and then factorize it into $(2,1)$ and
$(1,1,1)$ structures to obtain enough curves and degrees of
freedom to construct models with minimum MSSM matter contents. The
restrictions of the line bundles associated with two $U(1)$ gauge
groups to the matter curves can modify the chirality of matter
localized on the curves. This modification generally results in
plenty of exotic fields that may cause troubles in the
phenomenological interpretation of the models.

One way to arrange the matter content in the conventional $E_6$
GUT model building is that all the MSSM matter and Higgs fields
are included in the same $\bf 27$-plet with three copies and the
Yukawa coupling is ${\bf 27}\cdot{\bf 27}\cdot{\bf 27}$. Such kind
of interaction implies a structure of either one curve
intersecting itself twice or three curves intersecting, which
causes difficulties in geometry or the mass hierarchy structure in
F-theory model building. Therefore, we adopt an alternate way that
the weak scale Higgs particles are assigned to another $\bf 27$
curve while the representations of their original assignments in
the matter $\bf 27$ curve are taken as exotic leptons. By
additional symmetries such as baryon and lepton numbers, we can
rule out the undesired interactions coupled to the exotic fields.
The $(2,1)$ factorization providing two curves ${\bf 27}^{(a)}$
and ${\bf 27}^{(b)}$ with the interaction ${\bf 27}^{(a)}\cdot{\bf
27}^{(a)}\cdot{\bf 27}^{(b)}$ satisfies the basic requirements of
this picture. One the other hand, the $(1,1,1)$ factorization
confines three curves to the interaction ${\bf 27}^{(l_1)}\cdot
{\bf 27}^{(l_2)}\cdot{\bf 27}^{(l_3)}$. In this case we have to
distribute the MSSM matter to both ${\bf 27}^{(l_1)}$ and ${\bf
27}^{(l_2)}$ curves while the electroweak Higgs fields are
assigned on the third curve. The fermion mass matrices are
generally not able to admit the hierarchical structures except
they are tuned by appropriate flux restrictions. As mentioned
before, the additional one or more $({\bf 27}+\overline{\bf 27})$
pairs can be included to make sure that the gauge unification
occurs. These vector-like pairs generically exist on the curves in
F-theory and can be assigned to the same curve containing the
electroweak Higgs fields. However, the exact number of the
vector-like pairs on a matter curve is still unclear in the
present construction, so we assume that there exits at least one
pair.

We demonstrate several models both in the $(2,1)$ and $(1,1,1)$
factorizations with geometric backgrounds $dP_2$ and $dP_7$
studied in \cite{Caltech:global01} and
\cite{Blumenhagen:global02}, respectively. We also discuss the
cases that the restrictions of the line bundles associated with
$U(1)$s to the curves are trivial or non-trivial. Due to the
chirality constraints to the fields on the bulk, it is hard to
construct consistent $U(1)$ fluxes in $dP_2$. Therefore, we only
demonstrate a three-family $E_6$ GUT model without gauge breaking
in the $dP_2$ geometry. On the other hand, the $dP_7$ geometry has
more degrees of freedom for the parameters to build realistic
models. We therefore show in the $(2,1)$ case an example of a
three-generation model without $U(1)$ flux restrictions, and an
example with non-trivial $U(1)$ flux restrictions which gives rise
to exotic particles. In the $(1,1,1)$ factorization, we also
present an example of three-family model without flux restriction.
In that case there are two flavors on one matter curve and the
third flavor on the other. In the model with non-trivial flux
restrictions, we adjust the parameters so that the total chirality
of each representation on the two matter curves remain three while
the hierarchies of the mass matrices can be maintained. Regardless
of the exotic fields, the matter contents of our examples are
conventional and the corresponding phenomenology has been
discussed in the literature. Giving an appropriate interpretation
for the exotic fields remains a challenge in the semi-local/global
F-theory model building.

There are several interesting subgroups of the $E_6$ gauge group
and we only discuss the rank 5 scenario in this paper.  It would
be interesting to construct rank 6 models with $U(1)^3$ fluxes, as
well as the Pati-Salam-like and trinification-like models with
appropriate non-abelian gauge fluxes in F-theory.  We leave these
possibilities for future work.

\renewcommand{\thesection}{}
\section{\hspace{-1cm} Acknowledgments}

The work of CMC is supported in part by the Austrian Research
Funds FWF under grant I192. The work of YCC is supported in part
by the NSF under grant PHY-0555575 and by Texas A\&M University.

\vspace{.2 in}

\appendix{\bf\Large \hspace{-1cm}}

\section{Breaking via $E_6\ra SU(6)\times SU(2)$}

We list other possibilities of the subgroups after breaking $E_6$
by the $SU(2)\times U(1)^2$ flux.  The full matter content of $\bf
27$ and the corresponding $U(1)$ charges are presented.

\paragraph{Case 1. $SU(6)\ra SU(5)\times U(1)$}

\begin{equation}
\begin{array}{ccl}
E_6 & \xrightarrow[SU(2)] & SU(6)\times [SU(2)] \\
& \xrightarrow[U(1)_c] & SU(5)\times [SU(2)\times
U(1)_c] \\
& \xrightarrow[U(1)_d] & SU(3)\times SU(2)\times [SU(2)\times
U(1)_c\times U(1)_d] \\ \\
{\bf 27} & \rightarrow & {\bf (\bar 6,2)} + {\bf (15,1)} \\
& \rightarrow &  {\bf (\bar 5,2)}_{-1}+ {\bf (1,2)}_{5} +
{\bf (10,1)}_{2} + {\bf (5,1)}_{-4} \\
& \rightarrow &  {\bf (\bar 3,1,2)}_{-1,2}+{\bf (1,\bar
2,2)}_{-1,-3} +
{\bf (1,1,2)}_{5,0} \\
&& + {\bf (3,2,1)}_{2,1}+{\bf (\bar 3,1,1)}_{2,-4} + {\bf
(1,1,1)}_{2,6} +{\bf (3,1,1)}_{-4,-2}+{\bf (1,2,1)}_{-4,3}
\end{array}
\end{equation}
\begin{equation}
U(1)_c=\frac{1}{2}U(1)_a-\frac{3}{2}U(1)_b,~~
U(1)_d=\frac{3}{2}U(1)_a+\frac{1}{2}U(1)_b.
\end{equation}

\paragraph{Case 2. $SU(6)\ra SU(4)\times SU(2)\times U(1)$}

\begin{equation}
\begin{array}{ccl}
E_6 & \xrightarrow[SU(2)] & SU(6)\times [SU(2)] \\
& \xrightarrow[U(1)_e] & SU(4)\times SU(2)\times [SU(2)\times
U(1)_e] \\
& \xrightarrow[U(1)_f] & SU(3)\times SU(2)\times [SU(2)\times
U(1)_e\times U(1)_f] \\ \\
{\bf 27} & \rightarrow & {\bf (\bar 6,2)} + {\bf (15,1)} \\
& \rightarrow &  {\bf (\bar 4,1,2)}_{1}+ {\bf (1,\bar 2,2)}_{-2} +
{\bf (6,1,1)}_{-2} + {\bf (4,2,1)}_1 +{\bf (1,1,1)}_{4} \\
& \rightarrow &  {\bf (\bar 3,1,2)}_{1,1}+{\bf (1,1,2)}_{1,-3} +
{\bf (1,\bar
2,2)}_{-2,0} + {\bf (3,1,1)}_{-2,2} + {\bf (\bar 3,1,1)}_{-2,-2} \\
&& + {\bf (3,2,1)}_{1,-1}+{\bf (1,2,1)}_{1,3} + {\bf
(1,1,1)}_{4,0}
\end{array}
\end{equation}
\begin{equation}
U(1)_e=U(1)_a, ~~U(1)_f=U(1)_b.
\end{equation}

\paragraph{Case 3. $SU(6)\ra SU(3)\times SU(3)\times U(1)$}

\begin{equation}
\begin{array}{ccl}
E_6 & \xrightarrow[SU(2)] & SU(6)\times [SU(2)] \\
& \xrightarrow[U(1)_g] & SU(3)\times SU(3)\times [SU(2)\times
U(1)_g] \\
& \xrightarrow[U(1)_h] & SU(3)\times SU(2)\times [SU(2)\times
U(1)_g\times U(1)_h] \\ \\
{\bf 27} & \rightarrow & {\bf (\bar 6,2)} + {\bf (15,1)} \\
& \rightarrow &  {\bf (\bar 3,1,2)}_{-1}+ {\bf (1,\bar 3,2)}_1 +
{\bf (3,3,1)}_0 +
{\bf (\bar 3,1,1)}_2 +{\bf (1,\bar 3,1)}_{-2} \\
& \rightarrow &  {\bf (\bar 3,1,2)}_{-1,0}+ {\bf (1,\bar
2,2)}_{1,-1} +
{\bf (1,1,2)}_{1,2}+ {\bf (3,2,1)}_{0,1} + {\bf (3,1,1)}_{0,-2} \\
&& + {\bf (\bar 3,1,1)}_{2,0} +{\bf (1,\bar 2,1)}_{-2,-1} + {\bf
(1,1,1)}_{-2,2}
\end{array}
\end{equation}
\begin{equation}
U(1)_g=-\frac{1}{2}U(1)_a-\frac{1}{2}U(1)_b,~~
U(1)_h=\frac{1}{2}U(1)_a-\frac{1}{2}U(1)_b.
\end{equation}

\section{Breaking via Trinification}

\begin{equation}
\begin{array}{ccl}
E_6& \xrightarrow[~~~~~~~~~~~~~] & SU(3)\times SU(3)\times SU(3) \\
& \xrightarrow[SU(2)\times U(1)] & SU(3)\times SU(2)\times
[SU(2)\times U(1)_i] \times U(1)_j \\ \\
{\bf 27} & \rightarrow & {\bf (3,2,1)}_{-1,0} +{\bf (3,1,1)}_{2,0} \\
& & +{\bf (\bar 3,1,2)}_{0,1} +{\bf (\bar3,1,1)}_{0,-2}\\
& & +{\bf(1,2,2)}_{1,-1} +{\bf (1,2,1)}_{1,2} + {\bf
(1,1,2)}_{-2,-1}+{\bf (1,1,1)}_{-2,2}
\end{array}
\end{equation}
\begin{equation}
U(1)_i=-\frac{1}{2}U(1)_a+\frac{1}{2}U(1)_b,~~
U(1)_j=\frac{1}{2}U(1)_a+\frac{1}{2}U(1)_b.
\end{equation}

\renewcommand{\theequation}{\thesection.\arabic{equation}}
\setcounter{equation}{0}


\end{document}